\documentstyle[preprint,aps]{revtex}	
\begin{document}
%
%
\preprint{$
\begin{array}{l}
\mbox{UB-HET-97-02}\\[-3mm]
\mbox{FERMILAB-Pub-97/221-T}\\[-3mm]
\mbox{July~1997} \\   [2.mm]
\end{array}
$}
\title{QED Radiative Corrections to $Z$ Boson Production and the Forward
Backward Asymmetry at Hadron Colliders}
\author{U.~Baur}
\address{
Department of Physics, State University of New York, Buffalo, NY 14260, USA}
\author{S.~Keller}
\address{
Fermi National Accelerator Laboratory, Batavia, IL 60510, USA}
\author{W.K.~Sakumoto}
\address{
Physics Department, University of Rochester, Rochester, NY 14627, USA}
\maketitle
\vskip -3.mm
\begin{abstract}
\baselineskip15.pt  
The ${\cal O}(\alpha)$ radiative corrections to
the process $p\,p\hskip-7pt\hbox{$^{^{(\!-\!)}}$} \rightarrow \gamma^*,
\, Z \rightarrow \ell^+ \ell^-$ ($\ell=e,\,\mu$) are calculated.
Factorizing the collinear
singularity associated with initial state photon bremsstrahlung into 
the parton distribution functions, we find that initial state
corrections have a much smaller effect than final state radiative 
corrections. Due to mass singular logarithmic terms associated
with photons emitted collinear with one of the final state leptons, QED
radiative corrections strongly affect the shape of the di-lepton 
invariant mass distribution, the lepton transverse momentum spectrum,
and the forward backward asymmetry, $A_{FB}$. They lead to a sizeable
shift in the $Z$ boson mass extracted from data, decrease the di-lepton cross
section by up to 10\%, and increase the integrated forward backward
asymmetry in the $Z$ peak region by about 7\% at the Tevatron. We also 
investigate how experimental lepton identification requirements modify 
the effect of the QED corrections, and study the prospects for a high
precision measurement of $\sin^2\theta^{lept}_{eff}$ using the forward
backward asymmetry at the Large Hadron Collider (LHC).
\end{abstract}
%
\newpage
%
%
\narrowtext

\section{INTRODUCTION}

Over the last few years, the Standard Model (SM) of electroweak
interactions has been successfully tested at the one-loop level.
Experiments at LEP
and the SLC~\cite{Renton} have determined the properties of the $Z$
boson with a precision of 0.1\% or better, and correctly predicted the
range of the top quark mass from loop corrections~\cite{Renton}. 
Currently, the $Z$
boson mass is known to $\pm 2.0$~MeV, whereas the uncertainty of the $W$
mass, $M_W$, is $\pm 80$~MeV~\cite{Lanc}. A precise measurement of
$M_W$ and the top quark mass, $m_{top}$, would make it possible to derive 
indirect constraints on the Higgs boson mass, $M_H$, via top quark and
Higgs boson electroweak radiative corrections to $M_W$~\cite{tom}. With 
a precision
of 30~MeV (10~MeV) for the $W$ mass, and 2~GeV for the top quark mass, 
$M_H$ can be predicted with an uncertainty of about 50\% 
(20\%)~\cite{Tev2000}. Comparison of these constraints on $M_H$ 
with the mass obtained from direct 
observation of the Higgs boson in future collider experiments will be an
important test of the SM.

A significant improvement in the $W$ mass uncertainty is expected in the
near future from measurements at LEP~II~\cite{LEPWmass} and the Fermilab
Tevatron $p\bar p$ collider~\cite{Tev2000}. The ultimate precision
expected for $M_W$ from the combined LEP~II experiments is approximately 
40~MeV~\cite{LEPWmass}. At the Tevatron, integrated luminosities of
order 1~fb$^{-1}$ are envisioned in the Main Injector Era, and one
expects to measure the $W$ mass with a precision of approximately
50~MeV~\cite{Tev2000} per experiment. The prospects for a precise 
measurement of $M_W$ would further improve if a significant upgrade in 
luminosity beyond the goal of the Main Injector could be realized. 
With recent advances in
accelerator technology~\cite{GPJ}, Tevatron collider luminosities of
order $10^{33}\,{\rm cm}^{-2}\,{\rm s}^{-1}$ may become a reality,
resulting in integrated luminosities of up to
10~fb$^{-1}$ per year. With a total integrated luminosity of
30~fb$^{-1}$, one can target a precision of the $W$ mass of 15~--
20~MeV~\cite{Tev2000}. A similar or better accuracy may also be reached 
at the Large Hadron Collider (LHC)~\cite{KW}.

The determination of the $W$ mass in a hadron collider environment
requires a simultaneous precision measurement of the $Z$ boson mass,
$M_Z$, and width, $\Gamma_Z$. When compared to the value measured at LEP, 
the two quantities help to accurately determine the energy scale and
resolution of the electromagnetic calorimeter, and to constrain the
muon momentum resolution~\cite{CDFWmass,D0Wmass}. 

Analogous to the $W$ mass, a very high precision measurement of the
effective weak mixing angle,
$\sin^2\theta^{lept}_{eff}$~\cite{alberto}, can be used to extract 
information on the Higgs boson mass~\cite{Tev2000,Fisher}. At hadron
colliders, the
effective weak mixing angle can be determined from the forward backward
asymmetry, $A_{FB}$, in di-lepton production in the vicinity of the $Z$
pole~\cite{AFBCDF}.

In order to measure $A_{FB}$ and the $Z$ boson mass with high
precision in a hadron collider environment, it is necessary to fully 
understand and control higher order QCD and electroweak corrections. A
complete calculation of the full ${\cal O}(\alpha)$ 
radiative corrections to $p\,p\hskip-7pt\hbox{$^{^{(\!-\!)
}}$} \rightarrow \gamma^*,\, Z \rightarrow \ell^+ \ell^-$ has not
been carried out yet. In a previous calculation, only the final state 
photonic corrections had been included~\cite{BK,RGW}, using an 
approximation in which the sum of the soft and virtual part is
indirectly estimated
from the inclusive ${\cal O}(\alpha^2)$ $Z\to\ell^+\ell^-(\gamma)$
width and the hard photon bremsstrahlung contribution.  

In this paper, we present a more complete calculation of the ${\cal 
O}(\alpha)$ QED corrections to $p\,p\hskip-7pt
\hbox{$^{^{(\!-\!)}}$} \rightarrow \gamma^*,\, Z \rightarrow \ell^+
\ell^-$. Real and virtual 
initial and final state corrections, as well as the interference 
between initial and final state corrections are included. Purely 
weak corrections are expected to be very small and are therefore
ignored. Our
calculation also takes into account the mass of the final state leptons,
which regularizes the collinear singularity associated with final state 
photon radiation. Both $Z$ and photon exchange diagrams with all 
$\gamma - Z$ interference effects are incorporated. The di-lepton 
invariant mass thus is {\sl not} restricted to the $Z$ peak region.
Low mass Drell-Yan production is of interest because of the sensitivity
to parton distribution functions (PDF's) at small $x$ values~\cite{CDFDY}. 
High mass lepton pairs and the forward backward asymmetry above 
the $Z$ peak~\cite{Rosner} can be used to search for additional neutral 
vector bosons, and to constrain their couplings~\cite{CDFZp,CDFAFB}.
Results from our calculation have been used in 
Ref.~\cite{CDFAFB} to compare experimental data with the SM prediction
for $A_{FB}$. 

To perform our calculation, we use the Monte Carlo method for
next-to-leading-order (NLO) calculations described in 
Ref.~\cite{NLOMC}. The matrix elements for radiative $Z$ production and 
decay are taken from Ref.~\cite{BaBe} and~\cite{BaZe}. 
With the Monte Carlo method, it
is easy to calculate a variety of observables simultaneously and to 
simulate detector response. Special care has to be
taken in calculating the radiative corrections associated with
photon radiation from the incoming quarks and antiquarks. In the parton
model, quarks are assumed to be massless, and initial state photon radiation
results in collinear singularities.
The singular terms are universal to all orders in perturbation
theory and can be removed by universal collinear
counterterms generated by `renormalizing' the parton distribution
functions~\cite{spies,RPS}, in complete analogy to gluon emission in
QCD. A calculation of QED corrections using definite,
non-zero, values for quark masses~\cite{ward} and not factorizing the
corresponding collinear logarithms leads to a considerable
overestimation of the effects of initial state photon corrections.
However, QED corrections to the evolution of the parton distribution 
functions are not included in our calculation; a complete fit 
of the PDF's including all QED effects is beyond the scope of this paper. 
The technical details of our calculation are described in Sec.~II. 

Numerical results for $p\bar p$ collisions at
$\sqrt{s}=1.8$~TeV are presented in Sec.~III.
Due to the mass singular logarithms associated with final state photon
bremsstrahlung in the limit where the photon is emitted collinear with
one of the charged leptons, the di-lepton invariant mass distribution is
strongly affected by QED corrections, in particular in the vicinity of
the $Z$ boson resonance. As a result, the value extracted for $M_Z$ from
data is shifted to a lower value. The amount of the shift depends on
the lepton mass, and the detector resolution~\cite{CDFWmass,D0Wmass}. 
QED radiative corrections also
significantly affect the $Z$ boson production cross section when cuts
are imposed, the transverse momentum distribution of the leptons, and the
forward backward asymmetry below the $Z$ pole. For di-lepton masses
between 50~GeV and 100~GeV, the final state ${\cal
O}(\alpha)$ QED corrections are larger than the ${\cal O}(\alpha_s)$ QCD
corrections.

In Sec.~III, using a simplified model of the CDF detector as an
example, we also investigate how the finite energy and momentum
resolution of realistic detectors affect the QED corrections. Electrons 
and photons which are almost
collinear are difficult to discriminate, and the momenta of the two
particles are thus recombined into an effective electron
momentum~\cite{CDFWmass,D0Wmass} if they traverse the same calorimeter
cell, or, alternatively, if their separation in the
pseudorapidity -- azimuthal angle plane is below a critical value. The
second procedure completely eliminates the mass singular logarithms.
With the first method, residual effects of these terms remain when both
particles are almost collinear, but hit different calorimeter cells. In
practice, the numerical difference between the two procedures is
moderate; in both cases the significance of the QED corrections is
considerably reduced. In contrast, photons which are almost collinear
with muons are rejected if they are too
energetic~\cite{CDFWmass} which results in residual logarithmic
corrections to observable quantities in $\mu^+\mu^-$ production.
Transverse momentum and rapidity cuts are found to affect the lepton 
pair invariant
mass distribution and forward backward asymmetry in a similar way 
at the Born level and at ${\cal O}(\alpha^3)$. 

Recently, it has been suggested~\cite{Fisher}, that an ultra precise
measurement of $\sin^2\theta^{lept}_{eff}$ may be possible at the LHC 
($pp$ collisions at $\sqrt{s}=14$~TeV~\cite{LHC}) in
the muon channel, using the forward backward asymmetry in the $Z$ peak
region. At the LHC, the forward backward asymmetry is significantly
reduced compared to the Tevatron because of the larger sea -- sea quark
parton flux. We find that the
sensitivity of $A_{FB}$ to the effective weak mixing angle strongly
depends on the rapidity range over which the leptons can be detected.
The forward backward asymmetry at the LHC, including ${\cal O}(\alpha)$
QED and ${\cal O}(\alpha_s)$ QCD corrections, is
studied in detail in Sec.~IV. Finally, our conclusions are presented
in Sec.~V.

\section{Method of Calculation}

The calculation presented here employs a
combination of analytic and Monte Carlo integration techniques. Details
of the method can be found in Ref.~\cite{NLOMC}. The calculation of
di-lepton production in hadronic collisions at ${\cal O}(\alpha^3)$
includes contributions from the square of the Born graphs, the
interference between the Born diagrams and the virtual one loop graphs,
and the square of the real emission diagrams which we adopt from 
Refs.~\cite{BaBe,BaZe}. The diagrams contributing to the ${\cal 
O}(\alpha)$ QED corrections can 
be separated into gauge invariant subsets corresponding to 
initial and final state corrections. The squared matrix
element for the real emission diagrams is then given by
\begin{equation}
|{\cal M}^{2\to 3}|^2=|{\cal M}_i^{2\to 3}|^2+2Re[{\cal M}_i^{2\to
3}({\cal M}_f^{2\to 3})^*]+|{\cal M}_f^{2\to 3}|^2.
\end{equation}
${\cal M}_i^{2\to 3}$ and ${\cal M}_f^{2\to 3}$ are the separately gauge
invariant matrix elements associated with initial and final state
radiation. 

The basic idea of the
method employed here is to isolate the soft and collinear singularities
associated with the real photon emission subprocesses by partitioning
phase space into soft, collinear, and finite regions. This is done by 
introducing theoretical soft and
collinear cutoff parameters, $\delta_s$ and $\delta_c$. Using 
dimensional regularization~\cite{DIMREG}, the soft and collinear
singularities are exposed as poles in  $\epsilon$ (the number of
space-time dimensions is $N = 4 - 2\epsilon$ with $\epsilon$ a small
number). In the soft and collinear regions the cross section 
is proportional to the Born cross section. The soft region is defined by
requiring that the photon energy in the $q\bar q$ center of 
mass frame, $E_\gamma$, is $E_\gamma<\delta_s
\sqrt{\hat s}/2$ ($\hat s$ denotes the squared parton center of mass
energy). We can then evaluate, in $N$ dimensions, the
$2\to 3$ diagrams using the soft photon approximation, where the photon
momentum is set to zero in the numerator, and integrate over the
soft region. The soft singularities originating from final state photon 
radiation cancel against the corresponding singularities from
the interference of Born and final state virtual corrections. Similarly,
the soft singularities associated with initial state photon emission
and interference effects between initial and final state radiation
cancel against the corresponding singularities originating from initial state
vertex corrections, and the $Z\gamma$ and $\gamma\gamma$ box diagrams,
respectively. The remainder is then evaluated via
Monte Carlo integration as part of the $2\to 2$ contribution. For 
$E_\gamma>\delta_s\sqrt{\hat s}/2$, the real photon emission diagrams 
are calculated in four dimensions~\cite{BaBe,BaZe} using standard three 
body phase space Monte Carlo integration techniques. 

The collinear singularity associated with photon radiation from the 
final state lepton line is regulated by the finite lepton mass. The 
collinear singularities originating from initial state photon
bremsstrahlung are universal to all orders of perturbation theory and
can be cancelled by universal collinear counterterms generated
by renormalizing the parton distribution functions~\cite{spies,RPS}, in 
complete analogy to gluon emission in QCD~\cite{AL}. They occur when 
the final state photon and the partons in the initial state are
collinear so that denominators of propagators such as 
\begin{eqnarray}
\hat t & = & -2p_{\bar q}\cdot p_\gamma
\end{eqnarray}
and
\begin{eqnarray}
\hat u & = & -2p_{q}\cdot p_\gamma
\end{eqnarray}
vanish. Here, $p_q$ ($p_{\bar q}$) denotes the quark (anti-quark), and
$p_\gamma$ the photon four momentum vector.
Only $|{\cal M}_i^{2\to 3}|^2$ is divergent in the collinear limit;
the initial -- final state interference term, $Re[{\cal M}_i^{2\to 3}({\cal
M}_f^{2\to 3})^*]$ exhibits only soft singularities for massive final 
state leptons. In the collinear region, $|\hat t|,\,|\hat u|<
\delta_c\hat s$, $|{\cal M}_i^{2\to 3}|^2$ is evaluated in the 
leading pole
approximation. After $N$-dimensional integration over the photon phase
space variables, the explicit singularity can be factorized into the 
parton distribution functions. The remainder is evaluated as part of the
$2\to 2$ contribution. If $|\hat t|,\,|\hat u|>\delta_c\hat s$, the
$2\to 3$ diagrams are again evaluated numerically in four dimensions
using the full three body phase space. 

In order to treat the ${\cal O}(\alpha)$ initial state QED
corrections to di-lepton production in hadronic collisions in a
consistent way, QED corrections should be incorporated in the global 
fitting of the PDF's. 
Current fits~\cite{PDF} to the PDF's do not include QED corrections. A
study of the effect of QED corrections on the evolution of the parton 
distribution functions indicates~\cite{spies} that the modification 
of the PDF's is small. We have not attempted to include QED
corrections to the PDF evolution in the calculation presented here. The 
missing QED corrections to the PDF introduce an uncertainty which,
however, probably is much smaller than the present uncertainties on 
the parton distribution functions. 

Absorbing the collinear singularity into the PDF's introduces a
QED factorization scheme dependence. The squared matrix elements for different
QED factorization schemes differ by the finite ${\cal O}(\alpha)$ terms
which are absorbed into the PDF's in addition to the singular terms. 
As long as QED corrections to the PDF evolution are not
included, the ${\cal O}(\alpha^3)$ cross section will depend on the 
QED factorization scheme used. We have performed our calculation in the
QED ${\rm \overline{MS}}$ and DIS schemes, which are defined
analogously to the usual ${\rm \overline{MS}}$~\cite{MSBAR} and 
DIS~\cite{OWENSTUNG} schemes used in QCD calculations. Unless noted
otherwise, we will use the QED DIS scheme which minimizes the
effect of the ${\cal O}(\alpha)$ QED corrections on the PDF by
requiring the same expression for the leading and next-to-leading order 
structure function $F_2$ in deep inelastic scattering.

The $2\to 2$ contribution associated  with initial state radiative (ISR)
corrections, including the correction terms originating
from the absorption of the initial state collinear singularity, 
can be obtained from the
corresponding ${\cal O}(\alpha_s)$ QCD corrections~\cite{BR} by
replacing $(4/3)\alpha_s$ by $\alpha Q_q^2$, where $Q_q$ is the electric 
charge of the quark in units of the proton charge, in all relevant 
matrix element and cross section formulae. The $2\to 2$ contribution
induced by
the soft and virtual final state radiative (FSR) corrections is given by:
\begin{equation}
\Delta|{\cal M}^{2\to 2}|_f^2 = |{\cal M}^{Born}|^2\left[2\,{\alpha\over\pi}
\left(\log{\hat s\over m^2_\ell}-1\right )\log(\delta_s)+2\,{\alpha\over\pi}
\left({3\over 4}\,\log{\hat s\over m^2_\ell}+{\pi^2\over 6}-1\right)
+{\cal O}(\delta_s)\right ]
\label{EQ:FSR} 
\end{equation}
where $m_\ell$ is the lepton mass and
\begin{equation}
{\cal M}^{Born} = {\cal M}^\gamma + {\cal M}^Z
\end{equation}
is the Born $q\bar q\to \gamma^*,\,
Z\to\ell^+\ell^-$ matrix element. Finally, the $2\to 2$ contribution
induced by the ${\cal O}(\alpha^3)$ initial -- final state interference
correction terms is given by
\begin{eqnarray}
\label{EQ:INT}
\Delta|{\cal M}^{2\to 2}|^2_{int} & = & -2Q_q\,{\alpha\over\pi}\,\beta_{int}
\log(\delta_s)\,|{\cal M}^\gamma|^2 \\ \nonumber
& & -2Q_q\,{\alpha\over\pi}\,\beta_{int}\,Re\left[{\cal M}^\gamma{\cal M}^{Z*}
\log\left({\hat s\delta_s^2\over M^2_Z-\hat
s-i\hat s\gamma_Z}\right )\right] \\[2.mm] \nonumber
& & -2Q_q\,{\alpha\over\pi}\,\beta_{int}\,\log\left\vert{\hat s\delta_s
\over M^2_Z-\hat s-i\hat s\gamma_Z}\right \vert |{\cal M}^Z|^2 
\\[2.mm] \nonumber
& & +~{\rm finite~\gamma\gamma~and~\gamma}Z~{\rm box~terms}
\end{eqnarray}
with
\begin{equation}
\beta_{int} = \log\left({\hat t_1\over\hat u_1}\right)
\end{equation}
and
\begin{equation}
\gamma_Z={\Gamma_Z\over M_Z}.
\end{equation}
In our calculation, we use the full $\hat s$ dependent width in the $Z$ 
boson propagator.
The $\hat t_1$ and $\hat u_1$ are Mandelstam variables of the $2\to 2$
reaction:
\begin{eqnarray}
\hat t_1 & = & -2p_q\cdot p_{\ell^+}, \\
\hat u_1 & = & -2p_q\cdot p_{\ell^-}.
\end{eqnarray}
The finite terms from the $\gamma\gamma$ and $\gamma Z$ box diagrams are
identical to those in $e^+e^-\to q\bar q$ and can be found in 
Refs.~\cite{Hollik} and~\cite{HLK}. 

\begin{figure}
\vskip 15cm
\includegraphics{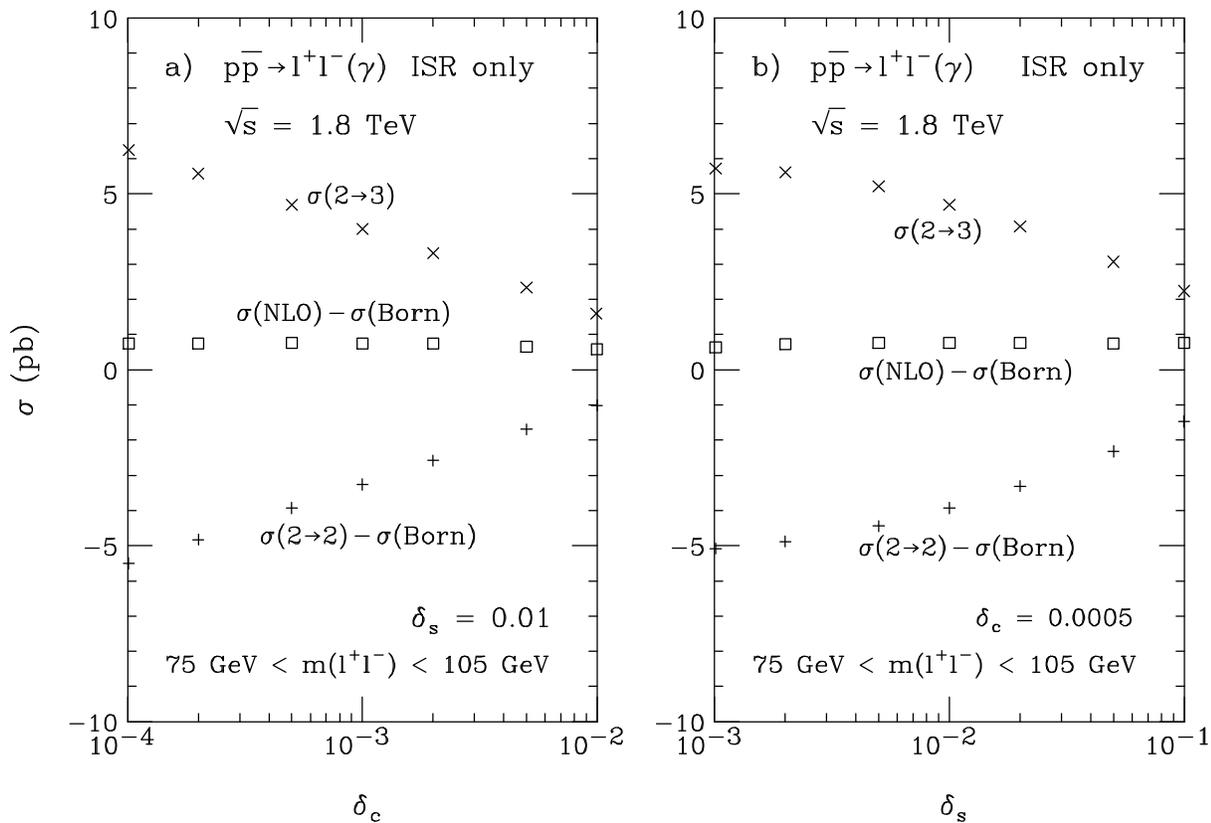}
\caption{The $p\bar p\to\ell^+\ell^-(\gamma)$, ($\ell=e,\,\mu$) cross 
section for $\protect{\sqrt{s}=1.8}$~TeV and
$75~{\rm GeV}<m(\ell^+\ell^-)$ $<105$~GeV as a function of a) $\delta_c$ for
$\delta_s=0.01$, and b) $\delta_s$ for $\delta_c=0.0005$, including
initial state radiation corrections only. Shown are $\sigma(2\to 2)
-\sigma({\rm Born})$, $\sigma(2\to 3)$, and $\sigma({\rm
NLO)}-\sigma({\rm Born})$. $\sigma({\rm NLO})$ denotes the ${\cal 
O}(\alpha^3)$ cross section.}
\label{FIG:ONE}
\end{figure}

\begin{figure}
\phantom{x}
\vskip 15cm
\includegraphics{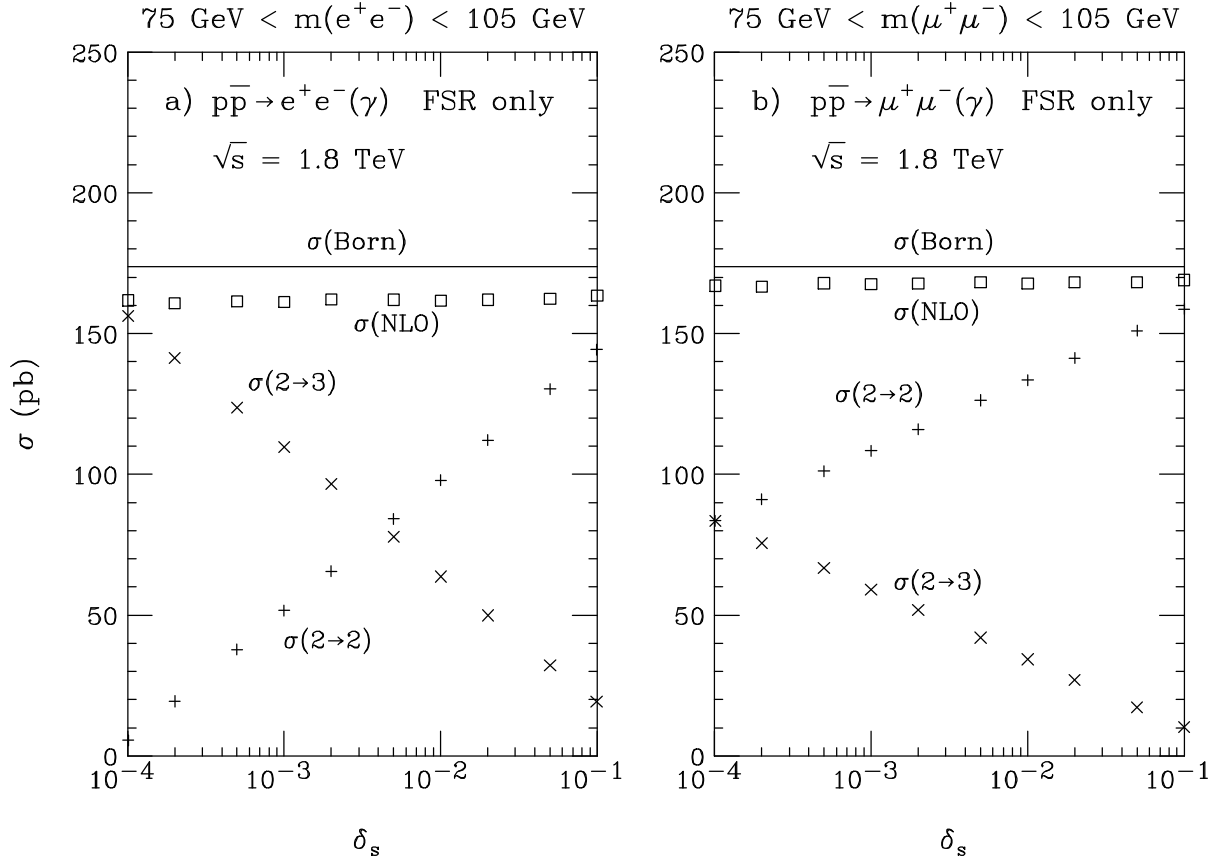}
\caption{The cross section a) $\sigma(p\bar p\to e^+e^-(\gamma))$
and b) $\sigma(p\bar p\to\mu^+\mu^-(\gamma))$ as a function of 
$\delta_s$, including final state radiation corrections only, for 
$\protect{\sqrt{s}=1.8}$~TeV  and $75~{\rm GeV}<m(\ell^+\ell^-)
<105$~GeV. Shown are the $2\to 2$ and $2\to 3$ contributions, and the
total ${\cal O}(\alpha^3)$ cross section. The solid line represents the
Born cross section.}
\label{FIG:TWO}
\end{figure}

The end result of the calculation consists of two sets of weighted
events corresponding to the $2\to 2$ and $2\to 3$ contributions.
Each set depends on the parameters $\delta_s$ and $\delta_c$. The sum of
the two contributions, however, must be independent of $\delta_s$
and $\delta_c$, as long as the two parameters are taken small
enough so that the approximations used are
valid. In Figs.~\ref{FIG:ONE} and~\ref{FIG:TWO} we show the dependence
of the $p\bar p\to\ell^+\ell^-(\gamma)$ cross section in the $Z$ peak
region ($75~{\rm GeV}<m(\ell^+\ell^-)<105$~GeV) on $\delta_s$ and 
$\delta_c$; $m(\ell^+\ell^-)$ denotes the di-lepton invariant mass.
To compute the cross section, we use here and in all subsequent figures 
the MRSA set of parton distribution functions~\cite{MRSA}, and take the
renormalization scale $\mu$ and the QED and QCD factorization scales,
$M_{QED}$ and $M_{QCD}$, to be $\mu^2=M_{QED}^2=M_{QCD}^2=\hat s$. 

Figure~\ref{FIG:ONE} displays the cross section as a function
of $\delta_c$ (Fig.~\ref{FIG:ONE}a) and $\delta_s$ (Fig.~\ref{FIG:ONE}b)
for initial state radiative corrections only. In order to 
exhibit the independence of the
cross section from the parameters $\delta_s$ and $\delta_c$ more
clearly, we have not included the Born cross section in the $2\to 2$
contribution.
The ISR corrections to the cross section for electron and muon final 
states are virtually identical. While the separate $2\to 2$ and $2\to 
3$ ${\cal O}(\alpha)$ 
contributions vary strongly with $\delta_s$ and $\delta_c$, the sum is
independent of the two parameters within the accuracy of the Monte Carlo
integration. The total contribution of initial state radiation diagrams
to the total cross section in the $Z$ pole region is found to be about
0.43\% of the Born cross section for the parameters chosen. In the QED
$\overline{\rm MS}$ scheme, the contribution of the ISR diagrams is 
about 10\% smaller than in the QED DIS scheme. QED corrections to the PDF's 
and purely weak one loop corrections to the matrix elements, both which 
are not included in our calculation, are expected to be of the same
order of magnitude. 

In Fig.~\ref{FIG:TWO}, we show the $p\bar p\to\ell^+\ell^-(\gamma)$
cross section in the $Z$ peak region ($75~{\rm GeV}<m(\ell^+\ell^-)<105$~GeV)
as a function of the soft cutoff parameter $\delta_s$ for
electron and muon final states for FSR corrections. 
Radiation of photons collinear with one of the leptons gives rise to
terms proportional to $\log(\hat s/m^2_\ell)\log(\delta_s)$ (see 
Eq.~(\ref{EQ:FSR})) in both the $2\to 2$ and $2\to 3$ contributions.
As demonstrated in Fig.~\ref{FIG:TWO}, these terms cancel
and the total cross section is independent of $\delta_s$. Due to the
smaller mass of the electron, the variation of the $2\to 2$ and $2\to 3$
contributions with $\delta_s$ is more pronounced in the electron case.
The solid line in Fig.~\ref{FIG:TWO} indicates the cross section in the
Born approximation. The total ${\cal O}(\alpha^3)$ cross section in the 
$e^+e^-(\gamma)$ ($\mu^+\mu^-(\gamma)$) case is found to be about 7\% 
(3\%) smaller than the Born cross section. The difference in the NLO 
$e^+e^-(\gamma)$ and $\mu^+\mu^-(\gamma)$ cross section 
can be traced to residual logarithmic correction terms which arise from 
the finite lepton pair invariant mass range considered in 
Fig.~\ref{FIG:TWO} (see Sec.~IIIA). If the integration would be carried 
out over the full
range $m(\ell^+\ell^-)>2m_\ell$, these terms would vanish~\cite{KLN}. From
Fig.~\ref{FIG:TWO} one also observes that, due to the residual
logarithmic terms, final state radiation effects are much larger than
those which originate from initial state radiation. 
The $2\to 2$ and the $2\to 3$ contributions to the FSR corrections each are
trivially independent of the collinear cutoff $\delta_c$. 

Similar to the FSR corrections, one can show that the sum of the
$2\to 2$ and $2\to 3$ contributions of the 
initial -- final state interference terms is independent of
$\delta_s$. The interference terms are typically of the same size as the
initial state corrections. 

The missing QED corrections to the PDF's create a dependence of the 
${\cal O}(\alpha)$ initial state corrections on the factorization scale
$M_{QED}$ which is stronger than that of lowest order calculation.
On the other hand, final state and initial 
-- final state interference terms depend on the factorization scale only
through the PDF's. These terms therefore exhibit a sensitivity 
to the factorization scale which is similar to that of the lowest order
calculation. Since the Born cross section and final state corrections
are much larger than corrections
from initial state radiation, the scale dependence of the
complete ${\cal O}(\alpha^3)$ cross section is similar to that of
the Born cross section. 

In conclusion, in Figs.~\ref{FIG:ONE} and~\ref{FIG:TWO} we demonstrated 
that the $p\,p\hskip-7pt\hbox{$^{^{(\!-\!)}}$} \rightarrow \gamma^*,
\, Z \rightarrow \ell^+ \ell^-(\gamma)$, cross section for
$75~{\rm GeV}<m(\ell^+\ell^-)<105$~GeV
is independent of the soft and collinear cutoff parameters $\delta_s$
and $\delta_c$ within the accuracy of the Monte Carlo integration. 
Independence of the cross section from these two
parameters can also be demonstrated for lepton pair invariant masses below
($m(\ell^+\ell^-)<75$~GeV) and above ($m(\ell^+\ell^-)>105$~GeV) the $Z$
peak. In the following, the soft and collinear 
cutoff parameters will be fixed to $\delta_s = 10^{-2}$ and $\delta_c = 
10^{-3}$, unless explicitly stated otherwise.

\section{${\cal O}(\alpha)$ Corrections to Di-lepton Production at the
Tevatron}

We shall now discuss the phenomenological implications of ${\cal
O}(\alpha)$ QED corrections to di-lepton production at the Tevatron
($p\bar p$ collisions at $\sqrt{s} = 1.8$~TeV). We first discuss the
impact of QED corrections on the lepton pair invariant mass
distribution and the forward backward asymmetry. We then
consider how the finite resolution of detectors and experimental lepton
identification requirements modify the effects of the QED corrections, 
and investigate how ${\cal O}(\alpha)$ QED corrections affect the
measured di-lepton ($Z$ boson) cross section within the cuts imposed. 
Finally, we study the effect of the full 
radiative corrections on the $Z$ boson mass extracted from data. 
The SM parameters used in our numerical simulations are $M_Z=91.
187$~GeV, $\alpha(M_Z^2)=1/128$, $\Gamma_Z=2.50$~GeV and 
$\sin^2\theta^{lept}_{eff}=0.2319$. These values are consistent with
recent measurements at LEP, SLC and the Tevatron~\cite{Renton}.

\subsection{QED Corrections to the Di-lepton Invariant Mass 
Distribution and $A_{FB}$} 

As we pointed out in Sec.~II, final state photon radiation leads to
corrections which are proportional to $\alpha\log(\hat s/m_\ell^2)$. These
terms are large, and are expected to significantly influence the 
shape of the di-lepton invariant mass distribution. The ${\cal 
O}(\alpha^3)$ $\ell^+\ell^-$ invariant mass distribution in the
vicinity of the $Z$ peak for the electron (solid line) and muon case
(dotted line) is shown in Fig.~\ref{FIG:THREE} together with the lowest 
order prediction (dashed line). No detector resolution effects or 
acceptance cuts are taken into account in any of the figures shown in this
subsection. QED corrections decrease (increase) the cross section at 
(below) the peak. At the peak position, the
differential cross section is reduced by a factor~\cite{Berends}
\begin{equation}
\rho\approx 1+\beta\log\left({\Gamma_Z\over M_Z}\right)
\end{equation}
with
\begin{equation}
\beta={2\alpha\over\pi}\left(\log{M_Z^2\over m_\ell^2}-1\right),
\label{EQ:BETA}
\end{equation}
{\it i.e.} by about 30\% in the electron case and by about 20\% in the 
muon case. The shape of the $Z$ boson resonance curve is seen to be 
considerably distorted by the ${\cal O}(\alpha)$ QED corrections. 
Photon radiation from one of the leptons lowers the di-lepton invariant
mass. Events from the $Z$ peak region therefore are shifted towards 
smaller values of $m(\ell^+\ell^-)$, thus reducing the
cross section in and above the peak region, and increasing the rate
below the $Z$ pole. Due to the $\log(\hat s/m_\ell^2)$ factor, the
effect of the corrections is larger in the electron case. The lowest 
order cross section is almost indistinguishable for the two lepton flavors. 

The size of the QED corrections to lepton pair production at the
Tevatron becomes more apparent in Fig.~\ref{FIG:FOUR} where we display
the ratio of the ${\cal O}(\alpha^3)$ and the Born cross section as a 
function of the lepton pair invariant mass. For
$40~{\rm GeV}<m(\ell^+\ell^-)<110$~GeV, the cross section ratio is seen
to vary rapidly. Below the
$Z$ peak, QED corrections enhance the cross section by up to a 
factor~2.7 (1.9) for electrons (muons). The maximum enhancement of the 
cross section occurs at $m(\ell^+\ell^-)\approx 75$~GeV. For 
$m(\ell^+\ell^-)<40$~GeV ($m(\ell^+\ell^-)>130$~GeV), ${\cal O}(\alpha)
$ QED corrections uniformly reduce the differential cross section
by about 7\% (12\%) in the electron case, and $\approx 2.5\%$
($\approx 7\%$) in the muon case. Integrating over the full di-lepton
invariant mass region, the large positive and negative corrections below and
above $M_Z$ cancel~\cite{KLN}. For 40~GeV$<m(\ell^+\ell^-)<M_Z$, a large
fraction of events contains a photon with energy $E_\gamma>1$~GeV.
As we have stated before, the dominant QED
radiative corrections are proportional to $\log(\hat s/m_\ell^2)$. The
$p\bar p\to\mu^+\mu^-$ cross section is therefore less affected by
radiative corrections than the $p\bar p\to e^+e^-$ rate. 

\begin{figure}
\phantom{x}
\vskip 15cm
\includegraphics{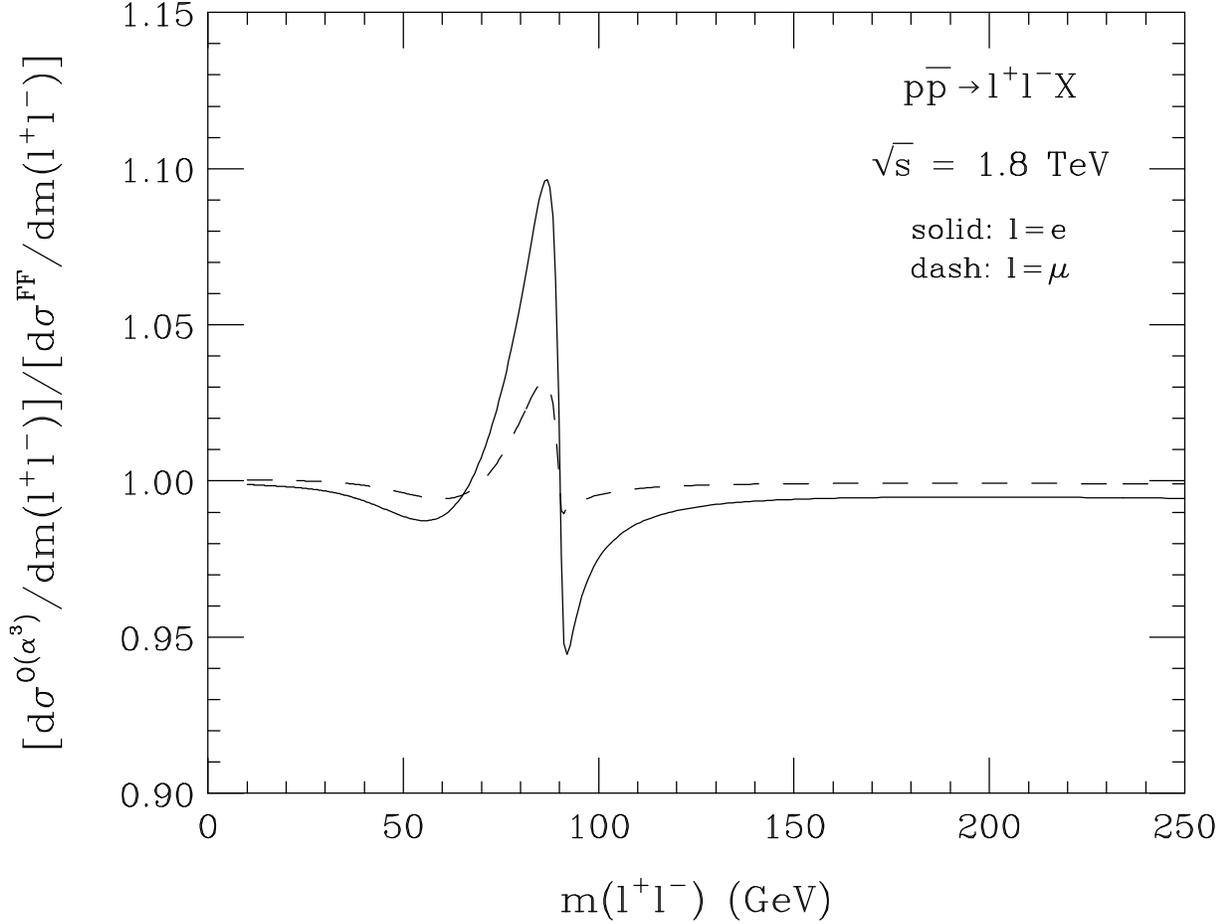}
\caption{Ratio of the \protect{${\cal O}(\alpha^3)$} cross section and
the cross section obtained in the fragmentation function approach
($\sigma^{FF}$) as a function of the di-lepton invariant
mass for $p\bar p\to\ell^+\ell^-X$ at 
$\protect{\sqrt{s}=1.8}$~TeV. The solid line shows the result obtained 
for final state electrons, whereas the dashed line displays the cross 
section ratio for muons. In the fragmentation function approach, only 
final state corrections are taken into account.}
\label{FIG:FOURA}
\end{figure}

It should be emphasized that the enhanced 
significance of the ${\cal O}(\alpha)$ QED corrections below the $Z$
peak is a direct consequence of the Breit-Wigner resonance of the $Z$
boson. The ${\cal O}(\alpha^2)$ radiative corrections therefore should
be a factor ${\cal O}((\alpha/\pi)\log(\hat s/m_\ell^2))$ smaller than the
${\cal O}(\alpha)$ corrections. The effect of higher order QED 
corrections on the $m(\ell^+\ell^-)$ distribution can be estimated using the
fragmentation function approach of Ref.~\cite{NT}. In this approach, the
radiatively corrected cross section is obtained by convoluting the
lowest order di-lepton cross section with a radiator function, which to
all order sums the dominant and non-dominant logarithmic terms.
Figure~\ref{FIG:FOURA} displays the ratio of the ${\cal O}(\alpha^3)$ cross
section and the cross section in the fragmentation function approach as
a function of $m(\ell^+\ell^-)$. Only final state corrections are taken
into account in the fragmentation function
approach. As for the ${\cal O}(\alpha)$ corrections, initial state
radiation contributions are expected to be small and, therefore, are
ignored. Figure~\ref{FIG:FOURA} shows that higher order final state 
QED corrections reduce the effect of the ${\cal O}(\alpha)$ corrections
and are indeed of the size naively expected. In the $Z$ peak
region, the higher order final state corrections vary rapidly with 
$m(\ell^+\ell^-)$ and change the 
differential cross section by up to 10\% (3\%) in the electron (muon) case. 

\begin{figure}
\phantom{x}
\vskip 15cm
\includegraphics{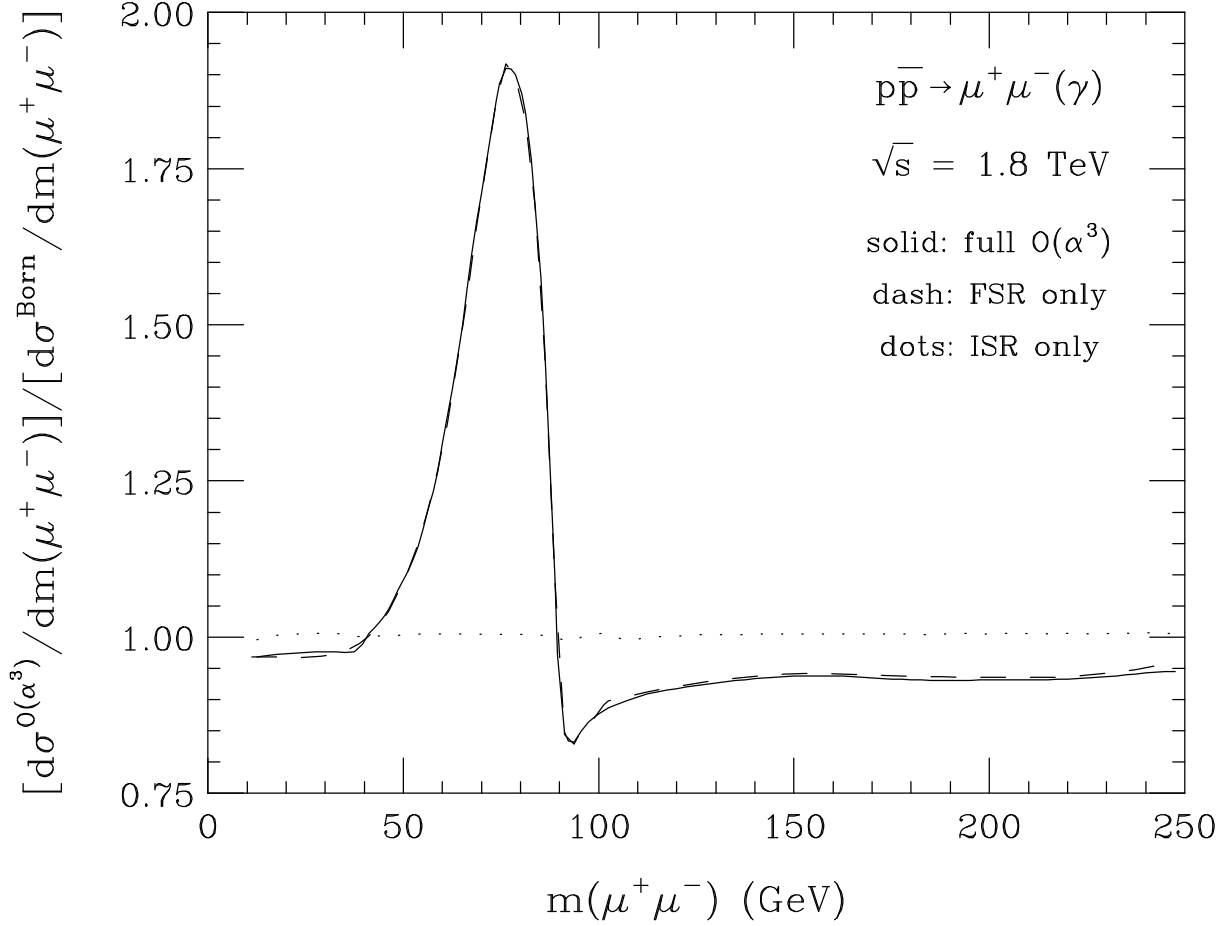}
\caption{Ratio of the \protect{${\cal O}(\alpha^3)$} and lowest order
differential cross sections as a function of the di-muon invariant mass 
for $p\bar p\to\mu^+\mu^-(\gamma)$ at 
$\protect{\sqrt{s}=1.8}$~TeV. The solid line gives the result for the
full set of \protect{${\cal O}(\alpha^3)$} QED diagrams. The dashed and
dotted lines show the ratio obtained taking only final state and 
initial state corrections, respectively, into account.}
\label{FIG:FIVE}
\end{figure}

In Fig.~\ref{FIG:FIVE}, we compare the impact of the full ${\cal 
O}(\alpha)$ QED corrections (solid line) on the muon pair invariant 
mass spectrum with that of final state (dashed line) and initial state
radiative corrections (dotted line) only. Qualitatively similar results 
are obtained
in the electron case. Final state radiative corrections are seen to
completely dominate over the entire mass range considered. They are
responsible for the strong modification of the 
di-lepton invariant mass distribution. In contrast, initial state
corrections are uniform and small ($\approx +0.4\%$). 

At small di-lepton invariant masses, photon exchange dominates and the
initial -- final state interference terms are almost completely
antisymmetric in $\cos\bar\theta^*$~\cite{Hollik,HLK,Bardin}, where 
$\bar\theta^*$ 
is the lepton scattering angle in the parton center of mass frame. The 
contribution of these interference terms
to the di-lepton invariant mass distribution is extremely small (0.01\% --
0.1\%) for $m(\ell^+\ell^-)<M_Z$. For values of $m(\ell^+\ell^-)$ sufficiently
above the $Z$ mass, initial -- final state interference terms reduce
the ${\cal O}(\alpha^3)$ cross section by about 1\%.

Next-to-leading order QCD corrections to lepton pair production in
$p\bar p$ collisions at Tevatron energies are known~\cite{BR} to
enhance the cross section by about 16\% -- 25\%. Since these are initial state
corrections, the NLO QCD to leading order cross section ratio varies only
slowly with the di-lepton invariant mass, similar to what we found for 
initial state QED corrections. Comparing the
size of the ${\cal O}(\alpha)$ QED and ${\cal O}(\alpha_s)$ QCD
corrections, one observes that they are of similar magnitude above the
$Z$ peak, but have opposite sign. In the invariant mass range
between 50~GeV and 100~GeV, QED corrections are significantly larger
than those induced by the strong interactions. The relative importance
of the QED corrections is due to the combined effect of mass 
singular logarithms associated with final state photon radiation, and
the $Z$ boson Breit-Wigner resonance. 

\begin{figure}
\phantom{x}
\vskip 15cm
\includegraphics{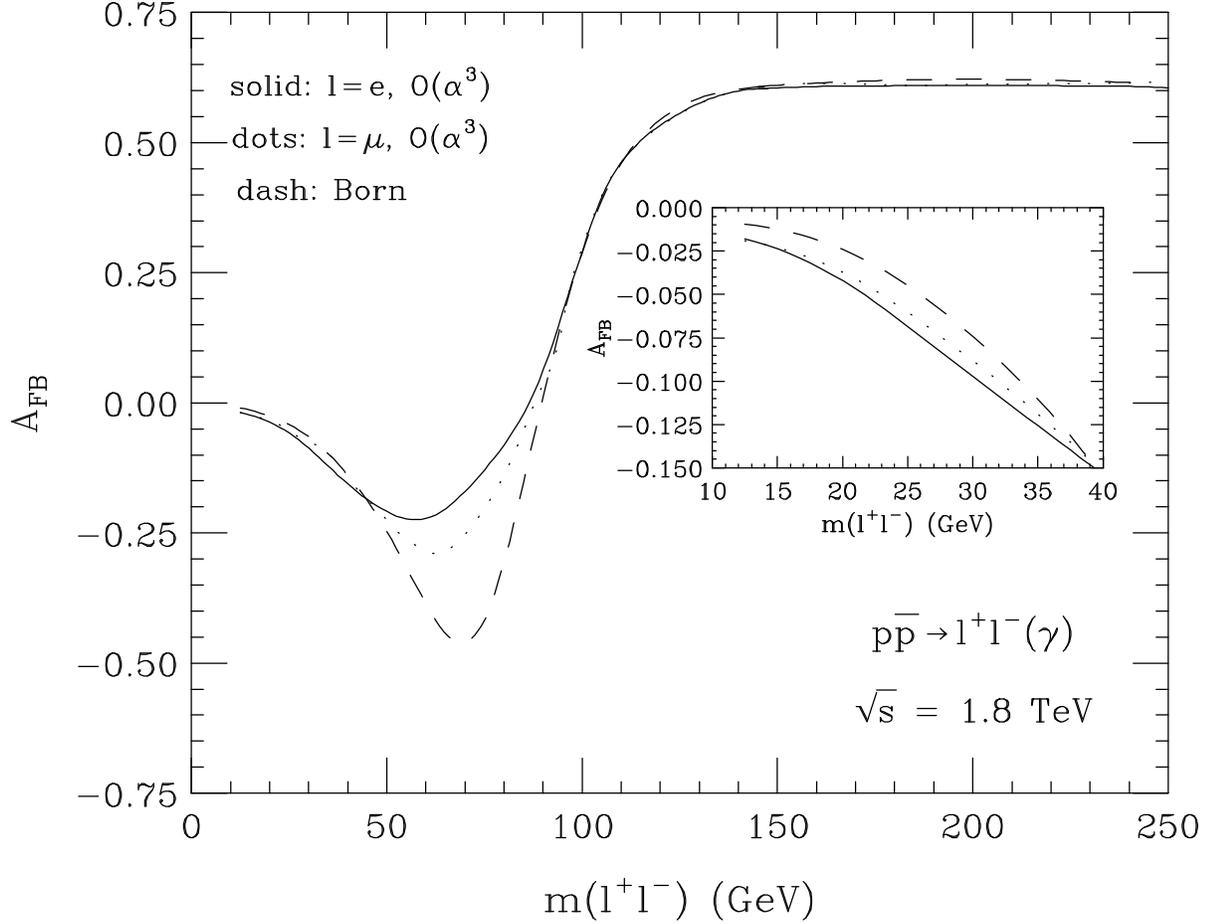}
\caption{The forward backward asymmetry, $A_{FB}$, as a function of the
di-lepton invariant mass for $p\bar p\to\ell^+\ell^-(\gamma)$ at
$\protect{\sqrt{s}=1.8}$~TeV. The solid and dotted lines show the
forward backward asymmetry including ${\cal O}(\alpha)$ QED corrections
for electrons and muons, respectively. The dashed line displays the
lowest order prediction of $A_{FB}$. The inset provides a closeup of
$A_{FB}$ in the low mass region. }
\label{FIG:SIX}
\end{figure}

Since QED corrections strongly affect the shape of the lepton pair 
invariant mass distribution below the $Z$ peak, one expects that they
may also have a significant impact on other observables in this region. In 
Fig.~\ref{FIG:SIX}, we show the forward backward asymmetry, $A_{FB}$, as
a function of the lepton pair invariant mass in the Born approximation (dashed
line), and including ${\cal O}(\alpha)$ QED corrections for electron
(solid line) and muon final states (dotted line). Here, $A_{FB}$ is 
defined by 
\begin{equation}
A_{FB}={F-B\over F+B}
\label{EQ:DEFAFB}
\end{equation}
where
\begin{equation}
F=\int_0^1{d\sigma\over d\cos\theta^*}\,d\cos\theta^*, \qquad
B=\int_{-1}^0{d\sigma\over d\cos\theta^*}\,d\cos\theta^*.
\label{EQ:DEFFB}
\end{equation}
$\cos\theta^*$ is given by~\cite{CDFAFB,CS}
\begin{equation}
\cos\theta^*={2\over m(\ell^+\ell^-)\sqrt{m^2(\ell^+\ell^-)
+p_T^2(\ell^+\ell^-)}}\left [p^+(\ell^-)p^-(\ell^+)-p^-(\ell^-)
p^+(\ell^+)\right ]
\label{EQ:CSTAR}
\end{equation}
with
\begin{equation}
p^\pm={1\over\sqrt{2}}\left (E\pm p_z\right ),
\end{equation}
where $E$ is the energy and $p_z$ is the longitudinal component of the
momentum vector. 
In this definition of $\cos\theta^*$, the polar axis is taken to be 
the bisector of the
proton beam momentum and the negative of the anti-proton beam momentum
when they are boosted into the $\ell^+\ell^-$ rest frame. In $p\bar p$ 
collisions at Tevatron energies, the flight direction of the incoming 
quark coincides with the proton beam direction for a large fraction of
the events. The definition of $\cos\theta^*$ in Eq.~(\ref{EQ:CSTAR})
has the advantage of minimizing the effects of the QCD corrections (see
below). In the limit of vanishing di-lepton $p_T$,
$\theta^*$ coincides with the angle between the lepton and the incoming 
proton in the $\ell^+\ell^-$ rest frame. 

Our result for $A_{FB}$ in the Born approximation agrees with that 
presented in Ref.~\cite{Rosner}. As expected, the ${\cal O}(\alpha)$ 
QED corrections to $A_{FB}$ are large in the region below the $Z$ peak. 
Since events from the $Z$ peak, where $A_{FB}$ is positive and small,
are shifted towards smaller values of $m(\ell^+\ell^-)$ by photon 
radiation, the forward backward asymmetry is significantly reduced in 
magnitude by radiative corrections for $50~{\rm 
GeV}<m(\ell^+\ell^-)<90$~GeV. 

The forward backward 
asymmetry in the Born approximation is small at low di-lepton masses, 
because of the dominance of photon exchange and the
vectorlike coupling of the photon to leptons.
For di-lepton masses below 40~GeV, the ${\cal O}(\alpha)$ initial --
final state interference correction terms are almost completely 
antisymmetric in $\cos\theta^*$ and comprise
the most important component of the QED corrections to $A_{FB}$. In this
region, the ${\cal O}(\alpha)$ QED 
corrections to $A_{FB}$ are therefore large. Initial -- final 
state interference terms do not contain any 
mass singular contributions. As a result, the
forward backward asymmetries for electron and muon final states
are similar for $m(\ell^+\ell^-)<40$~GeV. Details of the 
asymmetry in the low di-lepton mass region are
shown in the inset of Fig.~\ref{FIG:SIX}. Effects from purely weak
corrections are not included in our calculation. They could have a 
non-negligible effect on the forward backward asymmetry at low 
di-lepton masses, similar to the situation encountered in 
$e^+e^-\to\mu^+\mu^-$~\cite{BH1}.

In contrast to the lepton pair
invariant mass distribution, QED corrections to $A_{FB}$ are small for
$m(\ell^+\ell^-)>120$~GeV. They reduce the forward backward asymmetry by
about 1\% in this region. Initial and final state corrections to
$A_{FB}$ are of similar size for lepton pair invariant masses above the 
$Z$ peak. 

Recently, the CDF Collaboration has presented a first measurement of the
integrated forward backward asymmetry in $p\bar p\to e^+e^-X$ at the 
Tevatron for 
$m(e^+e^-)>105$~GeV, together with a more refined measurement in the
$Z$ peak region ($75~{\rm GeV}<m(e^+e^-)<105$~GeV)~\cite{CDFAFB}. In
Table~\ref{TAB:ONE}, we list the experimental values, together with the
theoretical prediction with and without ${\cal O}(\alpha)$ QED
corrections. QED corrections are seen to increase the 
asymmetry by about 8\% in the peak region. In the muon channel, the
increase in $A_{FB}$ for $75~{\rm GeV}<m(\mu^+\mu^-)<105$~GeV due to 
radiative corrections is approximately 4\%.

In the $Z$ peak region, $A_{FB}$ provides a tool to
measure $\sin^2\theta^{lept}_{eff}$~\cite{AFBCDF}. For
$75~{\rm GeV}<m(\ell^+\ell^-)<105$~GeV and $\sqrt{s}=1.8$~TeV, the forward
backward asymmetry can to a very good approximation be parameterized 
by~\cite{Rosner}
\begin{equation}
A_{FB}=b\left(a-\sin^2\theta^{lept}_{eff}\right)
\label{EQ:AFB}
\end{equation}
both in the Born approximation and including ${\cal O}(\alpha)$ QED 
corrections. For the parameters $a$ and $b$ we find in the Born 
approximation
\begin{equation}
a^{\rm Born}= 0.2454, \qquad\qquad b^{\rm Born}= 3.6
\label{EQ:AFBBORN}
\end{equation}
for $e^+e^-$ as well as $\mu^+\mu^-$ final states, and
\begin{equation}
a^{{\cal O}(\alpha^3)}=a^{\rm Born}+\Delta a^{\rm QED},\qquad b^{{\cal 
O}(\alpha^3)}=b^{\rm Born}+\Delta b^{\rm QED}
\label{EQ:AFBQED}
\end{equation}
with
\begin{equation}
\Delta a^{\rm QED}\approx 0.0010,\qquad\qquad \Delta b^{\rm QED}\approx 0
\end{equation}
for $p\bar p\to e^+e^-(\gamma)$, and
\begin{equation}
\Delta a^{\rm QED}\approx 0.0006,\qquad\qquad \Delta b^{\rm QED}\approx
-0.3
\end{equation}
for $p\bar p\to \mu^+\mu^-(\gamma)$. The change of the effective weak
mixing angle due to QED radiative corrections is a factor~3 to~4
larger than the current experimental uncertainty, 
$\delta\sin^2\theta^{lept}_{eff}=0.00024$~\cite{Renton}.

In Ref.~\cite{AFBCDF}, the approximation used to estimate the
electroweak corrections to $A_{FB}$ resulted in a significant dependence
of the correction to $\sin^2\theta^{lept}_{eff}$ on the infrared cutoff
used in the calculation. In contrast, as explained in detail in 
Sec.~II, our results are cutoff independent. This will make
it possible to substantially reduce the theoretical uncertainty of the
weak mixing angle extracted from future measurements of $A_{FB}$ at the
Tevatron. 

\subsection{Aspects of Experimental Lepton Identification and QED
Radiative Corrections}

It is well-known~\cite{KLN} that the mass singular logarithmic terms
which appear in higher orders of perturbation theory are 
eliminated when inclusive observables are considered. As explained
below, the finite
resolution of detectors prevents fully exclusive measurements.
Detector effects, which we have completely ignored so far, therefore
may significantly modify the effect of QED radiative corrections. To 
simulate detector acceptance, we impose the following
transverse momentum ($p_T$) and pseudo-rapidity ($\eta$) cuts:
\begin{quasitable}
\begin{tabular}{cc}
electrons & muons\\
\tableline
$p_{T}^{}(e)         > 20$~GeV  & $p_{T}^{}(\mu)         > 25$~GeV\\
$|\eta(e)|           < 2.4$     & $|\eta(\mu)|           < 1.0$\\
\end{tabular}
\end{quasitable}
In addition, we require that at least one electron (muon) is in the
central part of the detector: $|\eta(e)|<1.1$ ($|\eta(\mu)|<0.6$). 
These cuts approximately model the acceptance of the CDF detector for
electrons and muons. 
Uncertainties in the energy measurements of the charged leptons 
in the detector are simulated in the calculation by Gaussian smearing 
of the particle four-momentum vector with standard deviation $\sigma$
which depends on the particle type and the detector. The numerical results 
presented here were calculated using $\sigma$ values based on the 
CDF~\cite{RCDF} specifications. Similar results are obtained if the 
acceptances and energy resolutions of the D\O\ detector are 
used~\cite{D0Wmass}. 

The granularity of the detectors and the size of the electromagnetic
showers in the calorimeter make it difficult to discriminate between 
electrons and photons with a small opening angle. We therefore
recombine the four-momentum vectors of the electron 
and photon to an effective electron four-momentum vector if both
traverse the same calorimeter cell, assuming a calorimeter segmentation of
$\Delta\eta\times\Delta\phi=0.1\times 15^\circ$ ($\phi$ is the
azimuthal angle in the transverse plane). This procedure is similar to
that used by the CDF Collaboration. The segmentation chosen
corresponds to that of the central part of the CDF 
calorimeter~\cite{CDFWmass}. 
The D\O\ Collaboration uses a slightly different recombination procedure
where the electron and photon four-momentum vectors are combined if
their separation in the pseudorapidity -- azimuthal angle
plane, $\Delta R(e,\gamma)=\sqrt{(\Delta\eta(e,\gamma))^2+(\Delta\phi(e,
\gamma))^2}$, is smaller than a critical value, $R_c$. For 
$R_c=0.2$~\cite{D0Wmass}, the numerical results obtained 
are similar to those found with the calorimeter segmentation we use (see
above).

Muons are identified in a hadron collider detector by hits in the muon
chambers. In addition, one requires that the associated track is
consistent with a minimum ionizing particle. This limits the energy of a
photon which traverses the same calorimeter cell as the muon to be 
smaller than a critical value $E^\gamma_c$. In the subsequent
discussion, we assume $E^\gamma_c=2$~GeV~\cite{CDFWmass}.

\begin{figure}
\phantom{x}
\vskip 19cm
\includegraphics{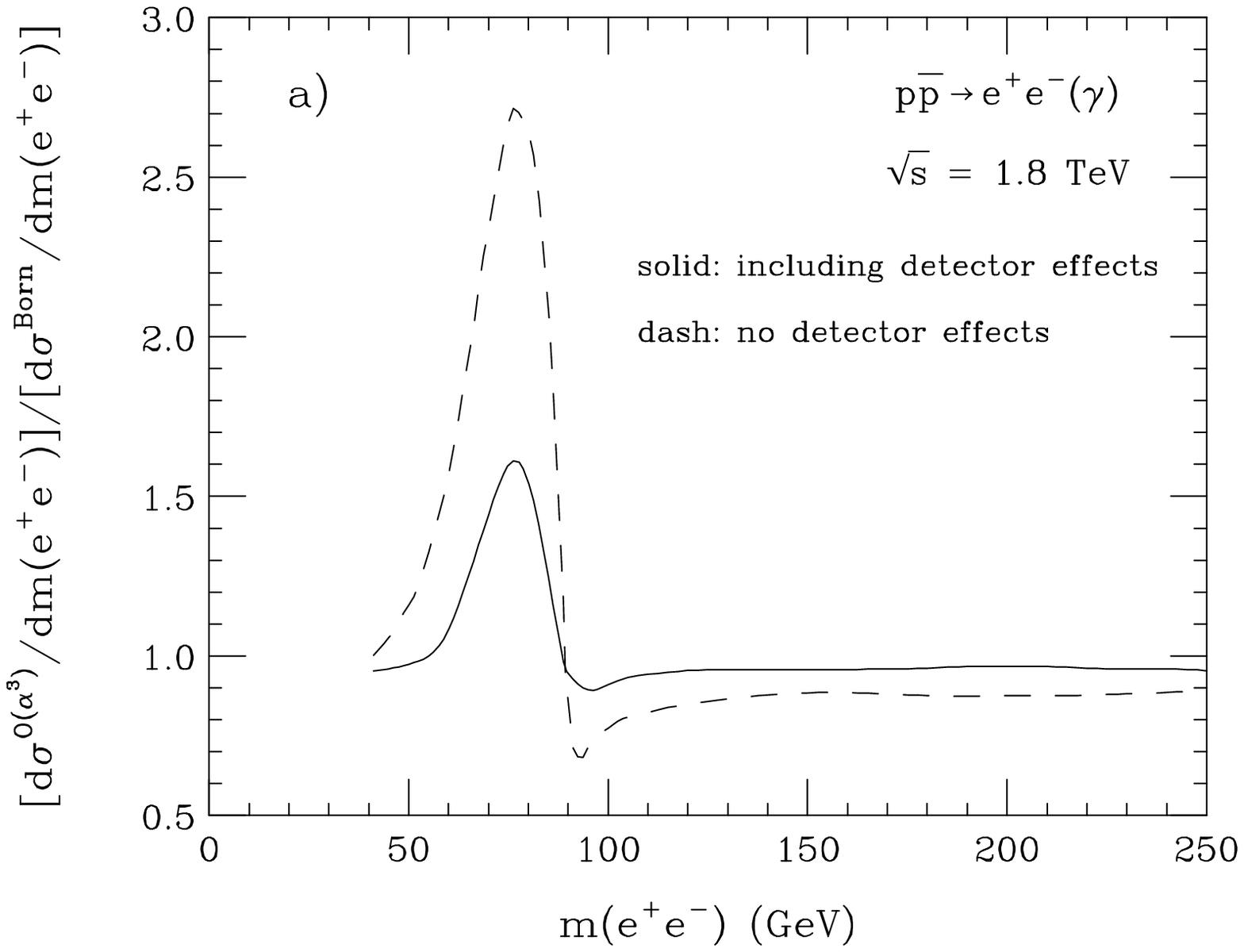}
\includegraphics{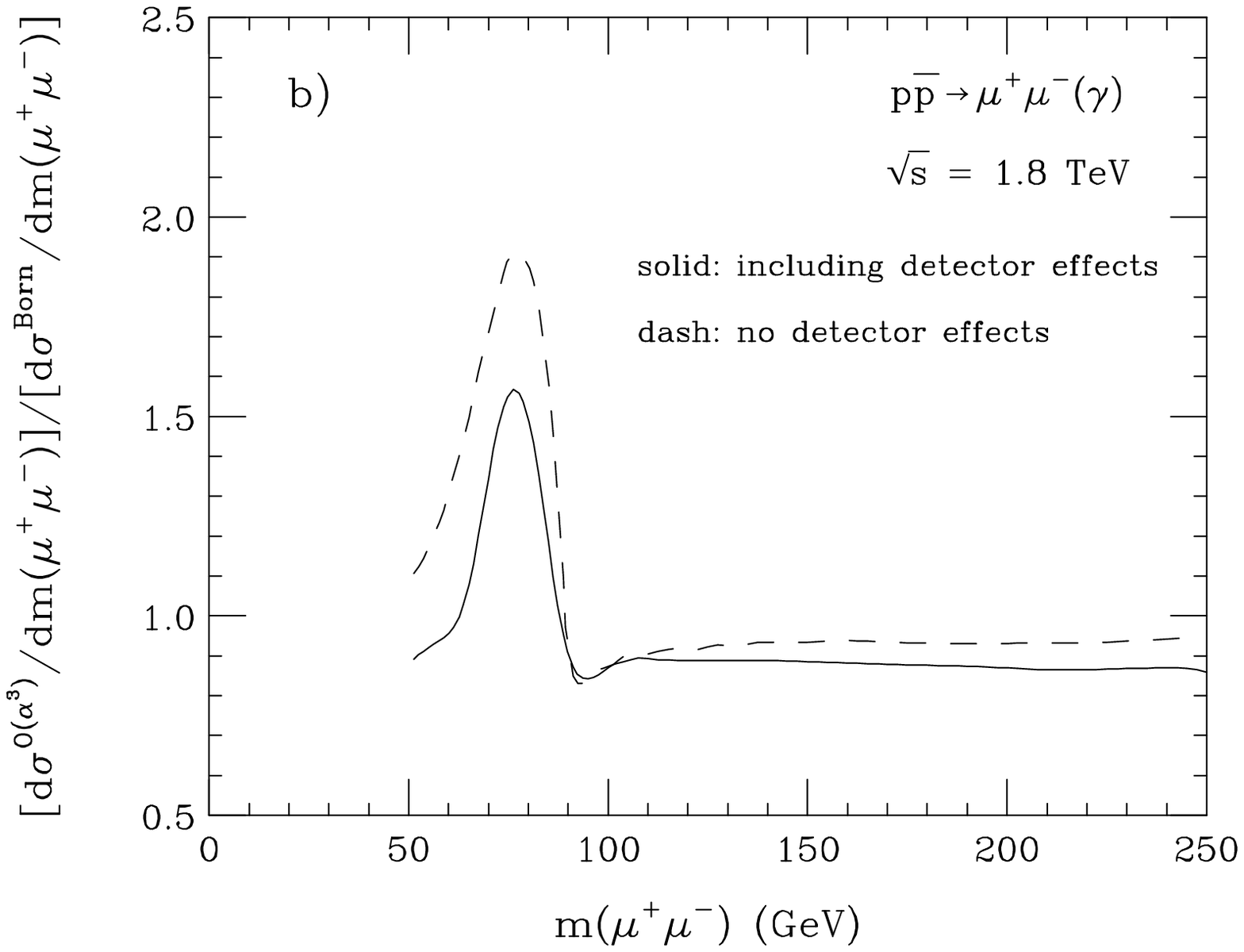}
\caption{Ratio of the \protect{${\cal O}(\alpha^3)$} and lowest order
differential cross sections as a function of the di-lepton invariant 
mass for a) 
$p\bar p\to e^+e^-(\gamma)$ and b) $p\bar p\to\mu^+\mu^-(\gamma)$ at 
$\protect{\sqrt{s}=1.8}$~TeV. The solid (dashed) lines show the cross
section ratio with (without) the detector effects described in the text. }
\label{FIG:SEVEN}
\end{figure}

In Fig.~\ref{FIG:SEVEN}a (Fig.~\ref{FIG:SEVEN}b) we show how detector
effects change the
ratio of the ${\cal O}(\alpha^3)$ to leading order differential cross
sections as a function of the $e^+e^-$ ($\mu^+\mu^-$) invariant mass. 
The finite energy resolution and the acceptance cuts have only a small
effect on the cross section ratio. The lepton identification criteria,
on the other hand, are found to have a large impact. Recombining the electron
and photon four-momentum vectors if they traverse the same calorimeter
cell greatly reduces the effect of the mass singular logarithmic terms. 
These terms survive only in the rare case when both
particles are almost collinear, but hit different calorimeter 
cells\footnote{In the case where the four-momentum vectors of the two
particles are recombined
for $\Delta R(e,\gamma)<R_c$, the mass singular terms are entirely
eliminated, and the lepton mass in the logarithmic terms is replaced by 
the minimum $e\gamma$ invariant mass.}. Although the recombination of the 
electron and photon momenta reduces effect of the ${\cal O}(\alpha)$ 
QED corrections, the remaining corrections are still sizeable. Below
(at) the $Z$ peak, they enhance (suppress) the lowest order
differential cross section by up to a factor~1.6 (0.9) [see 
Fig.~\ref{FIG:SEVEN}a]. For $m(e^+e^-)\gg M_Z$, the magnitude of the
QED corrections is reduced from approximately 12\% to 5\%. 

For muon final states (see Fig.~\ref{FIG:SEVEN}b), the requirement of 
$E_\gamma<E^\gamma_c=2$~GeV for a photon which traverses the same 
calorimeter cell as the muon reduces the hard photon part of the 
${\cal O}(\alpha^3)$ $\mu^+\mu^-(\gamma)$ cross section. As a result,
the magnitude of the QED corrections below the $Z$ peak is reduced. 
At the $Z$ pole the corrections remain unchanged, and
for $\mu^+\mu^-$ masses larger than $M_Z$ they become more pronounced. 
For $m(\mu^+\mu^-)>120$~GeV, QED corrections reduce the $\mu^+\mu^-$ 
cross section by 12\% to 14\%. 

We would like to emphasize that the survival of mass singular terms in
certain cases does not contradict the KLN theorem~\cite{KLN}. The KLN
theorem requires that mass singular logarithmic terms which appear in 
higher orders of perturbation theory are eliminated when inclusive 
observables are considered. Recombining the lepton and photon momenta
for small opening angles an inclusive quantity is formed, and
the mass singular logarithmic terms are eliminated in the reconstructed 
$\ell^+\ell^-$ invariant mass distribution. On the other hand, if the 
lepton and photon momenta are not combined, one performs an exclusive 
measurement, the KLN theorem does not apply, and logarithmic terms remain
present in the measured di-lepton invariant mass distribution.

It should be noted that the differential cross section ratio shown in
Fig.~\ref{FIG:SEVEN} becomes ill defined in the threshold region
$m(\ell^+\ell^-)\approx 2p_T^{cut}(\ell)$, where $p_T^{cut}(\ell)$ is
the charged lepton $p_T$ threshold. For $m(\ell^+\ell^-)\leq 
2p_T^{cut}(\ell)$, the Born cross section vanishes, and the cross section
ratio is undefined. The ${\cal O}(\alpha^3)$
cross section is small, but non-zero, in this region. The largest
contribution to the cross section for $m(\ell^+\ell^-)\leq 
2p_T^{cut}(\ell)$ originates from initial state radiation
configurations, where the leptons have a small relative opening angle
and are balanced by a high $p_T$ photon in the opposite hemisphere.
Close to the
threshold, $m(\ell^+\ell^-)\approx 2p_T^{cut}(\ell)$, large logarithmic
corrections are present, and for an accurate prediction of the cross
section those corrections need to be resummed. The results of 
Fig.~\ref{FIG:SEVEN} in this region should therefore be interpreted with
caution. Similar conclusions can also be drawn for the forward backward
asymmetry in the threshold region. 

\begin{figure}
\phantom{x}
\vskip 17cm
\includegraphics{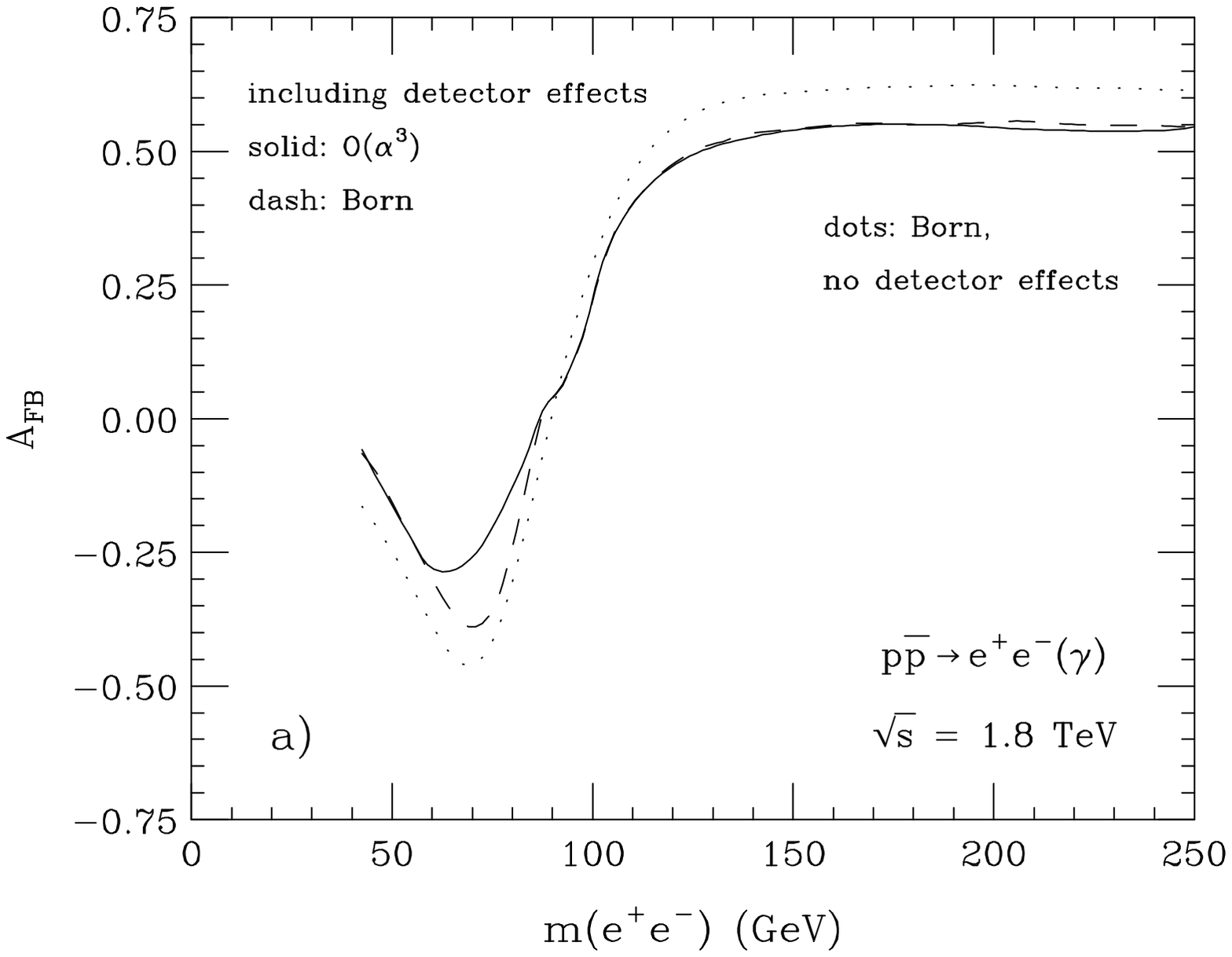}
\includegraphics{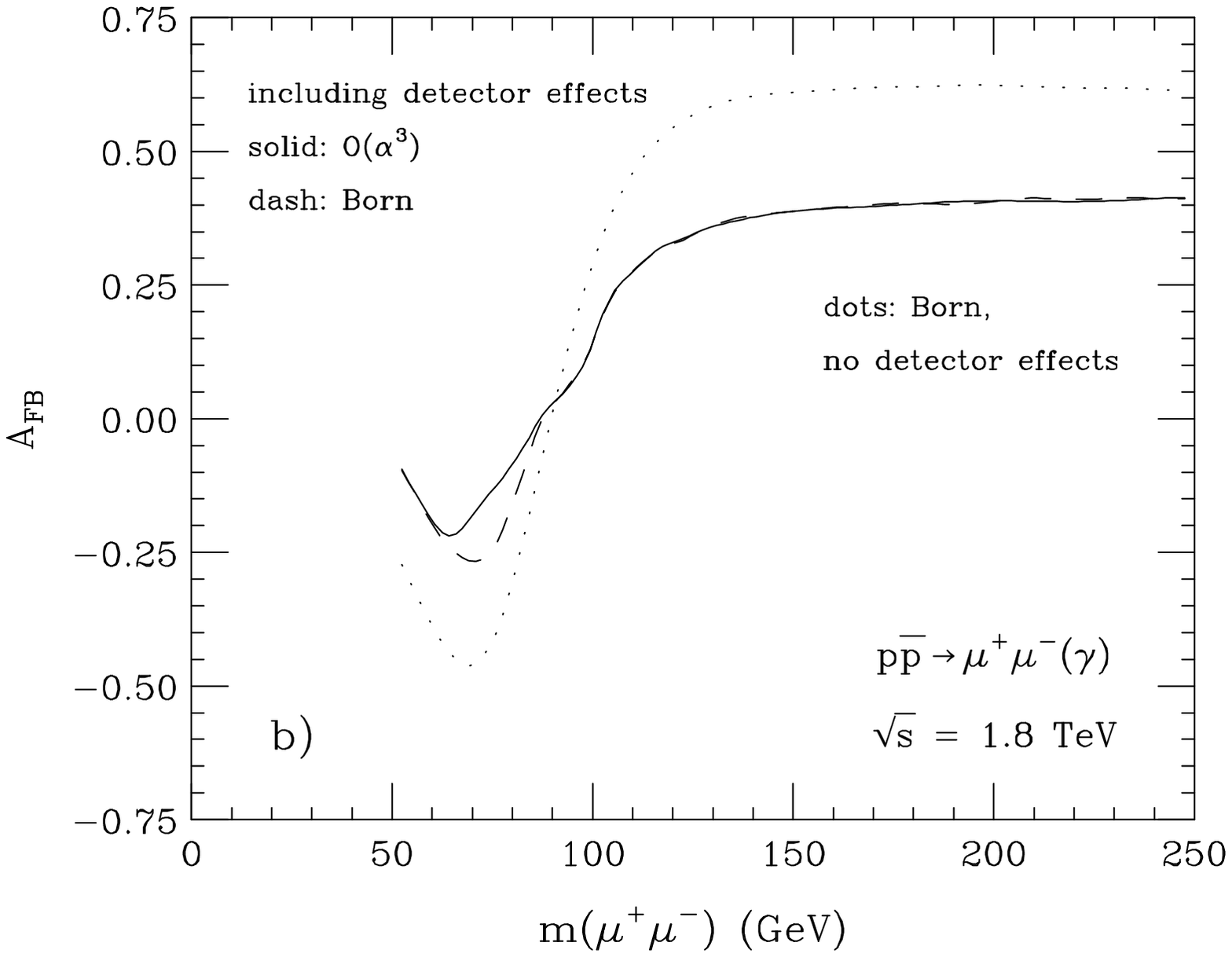}
\caption{The forward backward asymmetry, $A_{FB}$, a) for $p\bar p\to
e^+e^-(\gamma)$ and b) for $p\bar p\to\mu^+\mu^-(\gamma)$ at 
$\protect{\sqrt{s}=1.8}$~TeV as a function of the di-lepton invariant 
mass. The solid
lines show the result of the ${\cal O}(\alpha^3)$ calculation
including detector effects (see text for details). The dashed and dotted
lines represent the forward backward asymmetry in the Born
approximation with and without detector effects.}
\label{FIG:EIGHT}
\end{figure}

In Fig.~\ref{FIG:EIGHT}, we show how detector effects affect 
the forward backward asymmetry for electron (Fig.~\ref{FIG:EIGHT}a)
and muon final states (Fig.~\ref{FIG:EIGHT}b). In addition to the cuts
listed at the beginning of this subsection, we require~\cite{CDFAFB}
\begin{equation}
|\cos\theta^*|<0.8.
\end{equation}
For comparison, we also show the asymmetry in the Born approximation
without taking any detector related effects into account (dotted line). 
The finite lepton rapidity coverage and the $|\cos\theta^*|$ cut 
significantly reduce the forward backward asymmetry in
magnitude. Energy and momentum resolution effects 
broaden the $Z$ peak and thus introduce a characteristic $S$ type bending in
$A_{FB}$ at $m(\ell^+\ell^-)\approx M_Z$. Analogous to the di-lepton
invariant mass distribution, lepton identification
requirements substantially reduce the impact of QED radiative
corrections on the forward backward asymmetry below the $Z$ peak. 
For $m(\ell^+\ell^-) >100$~GeV, they have only a small effect on
$A_{FB}$, similar to the case where no detector effects are taken into
account. 

\begin{figure}
\phantom{x}
\vskip 15cm
\includegraphics{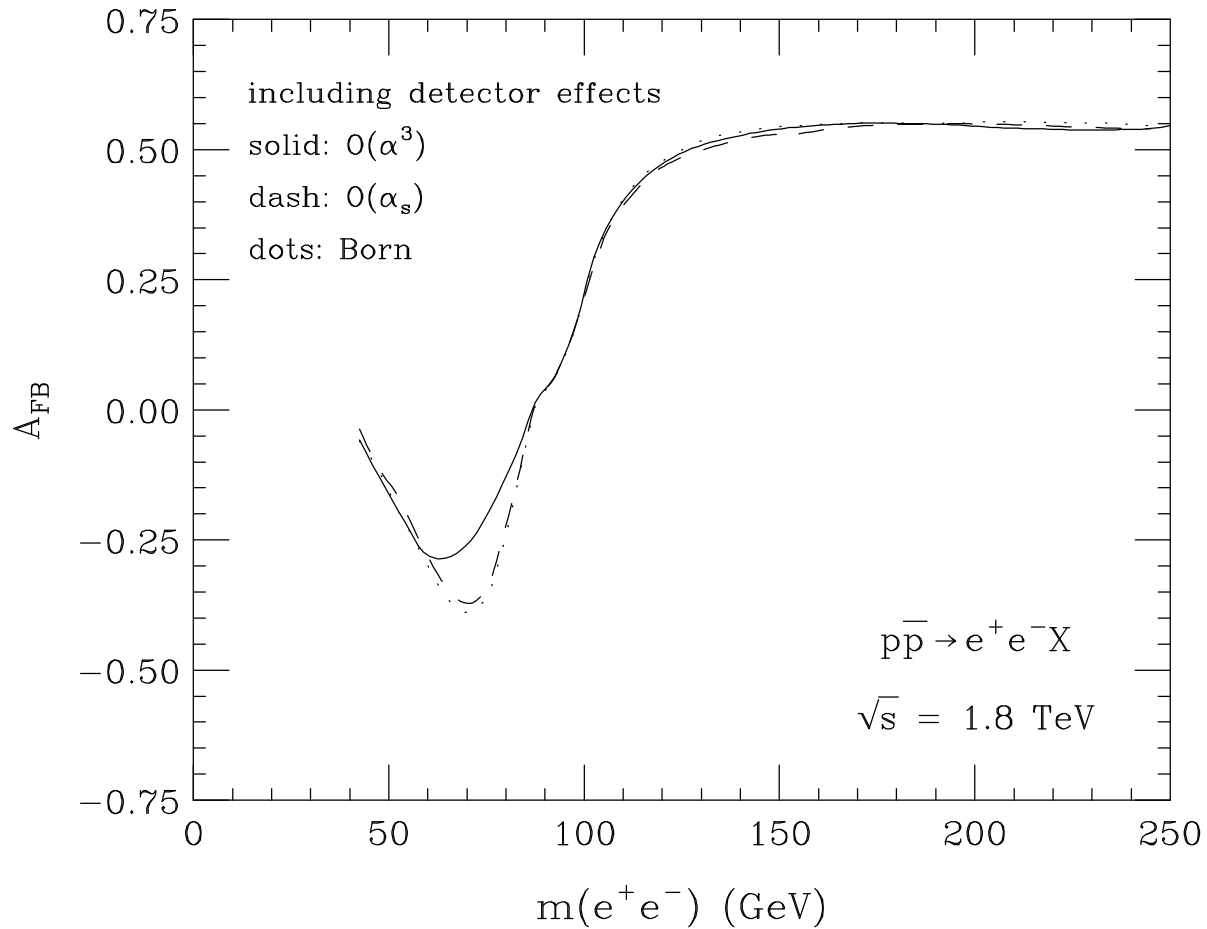}
\caption{The forward backward asymmetry, $A_{FB}$, including detector 
effects (see text for details) as a function of 
the $e^+e^-$ invariant mass for $p\bar p\to e^+e^-X$ at 
$\protect{\sqrt{s}=1.8}$~TeV. The curves are for the forward backward
asymmetry in the Born approximation (dotted line), including ${\cal
O}(\alpha)$ QED corrections (solid line), and including ${\cal
O}(\alpha_s)$ QCD corrections (dashed line). }
\label{FIG:NINE}
\end{figure}

Although QED corrections to the forward backward asymmetry are reduced 
in magnitude for $m(\ell^+\ell^-)<M_Z$ by 
experimental lepton detection and identification requirements, they are
still considerably larger than the NLO QCD corrections in this region.
This is demonstrated in Fig.~\ref{FIG:NINE} for the electron final
state. Similar results are obtained for $p\bar p\to\mu^+\mu^-(\gamma)$.
The ${\cal O}(\alpha_s)$ QCD corrections to $p\bar p\to Z,
\gamma^*\to\ell^+\ell^-X$ are calculated in the ${\rm \overline{MS}}$
scheme using the Monte Carlo approach of Ref.~\cite{NLOMC}. The
calculation generalizes that of Ref.~\cite{BR1} to include 
finite $Z$ width effects and virtual photon exchange diagrams. 
The QCD corrections to $A_{FB}$~\cite{CALLA} are found to be quite 
small. Below (above) the $Z$ peak, the magnitude of the forward
backward asymmetry is reduced by typically $\delta
A_{FB}/A_{FB}\approx -0.05$ ($\delta A_{FB}/A_{FB}\approx -0.02$). For 
$75~{\rm GeV}<m(\ell^+\ell^-)<105$~GeV, NLO QCD corrections decrease
the integrated asymmetry by $\delta A_{FB}/A_{FB}\approx -0.03$.
QED and QCD corrections to the integrated forward backward asymmetry in 
the $Z$ peak region have opposite signs.

\begin{figure}
\phantom{x}
\vskip 15cm
\includegraphics{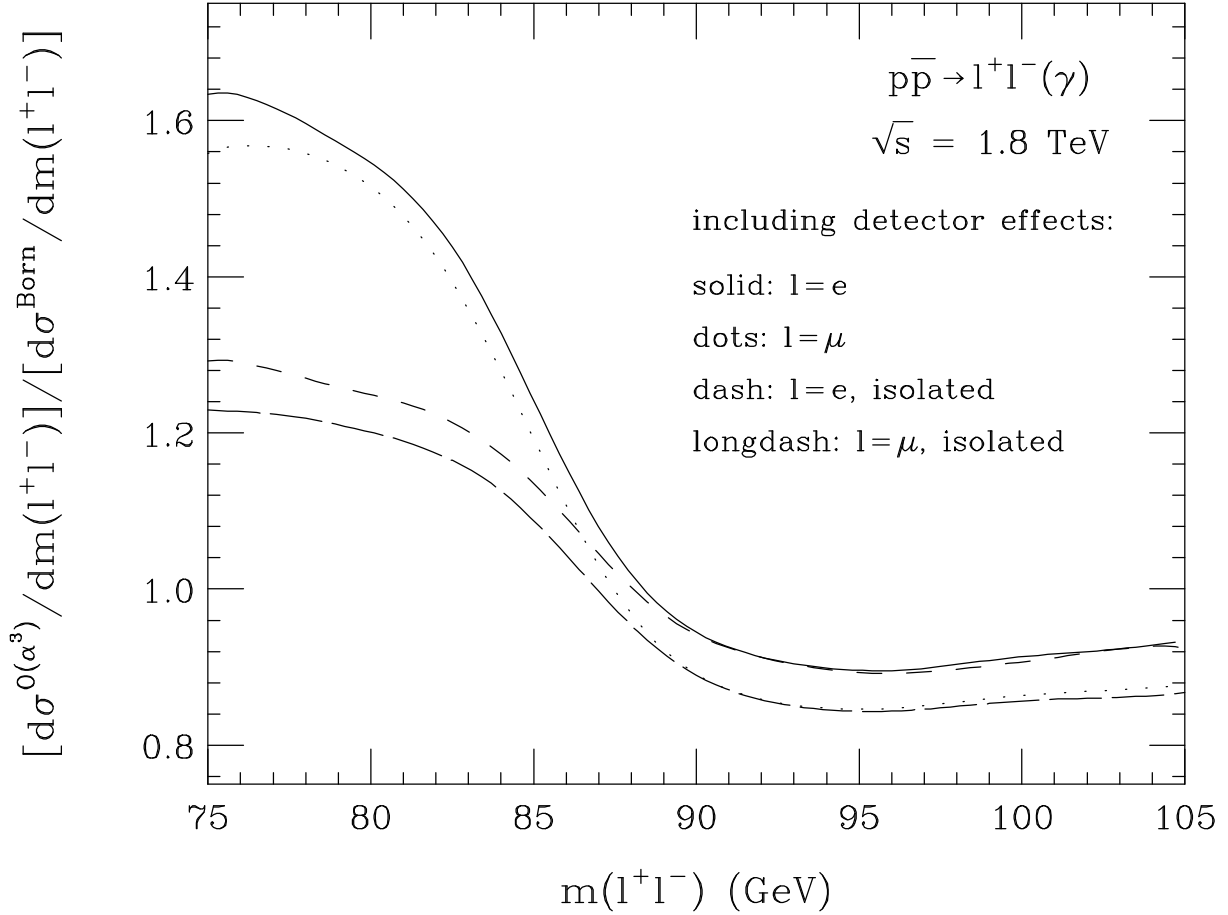}
\caption{Ratio of the \protect{${\cal O}(\alpha^3)$} and lowest order
differential cross sections, including detector effects (see text for
details), as a function of the di-lepton invariant 
mass for $p\bar
p\to\ell^+\ell^-(\gamma)$ at $\protect{\sqrt{s}=1.8}$~TeV in the $Z$
peak region. The solid and dotted lines show the 
cross section ratio without imposing a lepton isolation cut for
electrons and muons, respectively. The short-dashed and long-dashed 
lines give the result imposing in addition the isolation requirement of 
Eq.~(\protect{\ref{EQ:ISO}}) with $R_0=0.4$ and $\epsilon_E=0.1$.}
\label{FIG:TEN}
\end{figure}

To reduce the background from heavy flavor production processes, 
the leptons in $Z$ boson events are often required to be isolated. A lepton
isolation cut typically requires the transverse energy in a cone of
size $R_0$ about the direction of the lepton, $E^{R_0}_T$, to be 
less than a fraction, $\epsilon_E$, of the lepton transverse energy 
$E_T(\ell)$, {\it i.e.}
\begin{equation}
{E^{R_0}_T-E_T(\ell)\over E_T(\ell)}<\epsilon_E.
\label{EQ:ISO}
\end{equation}
Sometimes the energy, $E$, instead of the transverse energy is used in
the isolation requirement, Eq.~(\ref{EQ:ISO}). The isolation requirement
and the cut imposed on the photon energy in the muon case have similar 
effects. In Fig.~\ref{FIG:TEN}, we show how the lepton isolation 
requirement of Eq.~(\ref{EQ:ISO}) with $R_0=0.4$ and
$\epsilon_E=0.1$ modifies the effect of the ${\cal O}(\alpha)$ QED
corrections on the di-lepton invariant mass distribution in the $Z$ peak
region. The isolation cut is seen to mostly affect the mass region below
$M_Z$, reducing the maximum enhancement of the differential cross
section by QED radiative corrections from a factor $\sim 1.6$ to 1.2 --
1.3. In our calculation, for electrons, the isolation requirement is 
only imposed if the electron and photon are not recombined. 
${\cal O}(\alpha\alpha_s)$ corrections to di-lepton production are not
included in the results presented. These 
corrections are expected to increase $E_T^{R_0}$ somewhat, and 
therefore will modify the effect of the isolation cut.

In the past, the measurement of the $W$ and $Z$ boson cross sections has
provided a test of perturbative QCD~\cite{D0ZSIG,CDFZSIG,CDFZSIGN}.
With the large data set accumulated in the 1994-95 Tevatron collider 
run, the uncertainty associated with the integrated luminosity ($\approx 3.
6\%$~\cite{CDFZSIGN}) becomes a limiting factor in this measurement. This
suggests to use the measured $W$ and $Z$ boson cross sections to 
determine the integrated luminosity in future
experiments~\cite{CDFZSIGN,DITT1}. In order to accurately measure the 
integrated luminosity, it will be necessary not only to take the ${\cal 
O}(\alpha_s^2)$ corrections to the $W$ and $Z$ boson cross
sections into account, but also to correct for higher order QED effects. 

Experimentally, the $Z$ boson cross section is extracted from the
di-lepton cross section in a specified invariant mass interval around
the $Z$ boson mass, correcting for photon exchange and $\gamma Z$ 
interference effects. The size of the ${\cal O}(\alpha)$ QED
corrections to the total di-lepton cross section is sensitive to the 
lepton identification criteria, the acceptance cuts and the range of 
the di-lepton invariant masses selected (see Fig.~\ref{FIG:TEN}). 
In Table~\ref{TAB:TWO} we list the cross section ratio (``QED $K$-factor'')
\begin{equation}
K^{QED}={\sigma^{{\cal O}(\alpha^3)}\over\sigma^{\rm Born}}
\end{equation}
for $75~{\rm GeV}<m(\ell^+\ell^-)<105$~GeV ($\ell=e,\,\mu$). For 
comparison, we also tabulate the corresponding QCD $K$-factor,
\begin{equation}
K^{QCD}={\sigma^{{\cal O}(\alpha_s)}\over\sigma^{\rm Born}}~.
\end{equation}
One observes that the effect of the large QED corrections
found in $d\sigma/dm(\ell^+\ell^-)$ is strongly reduced when
integrating over a range in invariant mass which is approximately
centered at $M_Z$. Nevertheless, the QED corrections usually are 
not negligible when compared with the ${\cal O}(\alpha_s)$ QCD
corrections. QCD corrections enhance the $Z$ boson production rate,
whereas QED effects decrease the cross section for the 
invariant mass window chosen here. The total $p\bar p\to e^+e^-X$
($p\bar p\to\mu^+\mu^-X$) cross section is reduced by about 7\% (3\%)
by QED radiative corrections. As we have noted before, the dominant 
QED correction terms
are proportional to $\log(\hat s/m^2_\ell)$ in absence of detector
related effects. Without detector effects taken into account, QED 
corrections to $p\bar p\to e^+e^-X$ thus are larger than for di-muon 
production. The recombination of electron and photon momenta when the
opening angle between the two particles is small strongly reduces the 
effect of the QED corrections to the integrated
$e^+e^-$ cross section. In the muon case, lepton identification
requirements increase the magnitude of the QED corrections, and they 
almost compensate the cross section enhancement originating from ${\cal 
O}(\alpha_s)$ QCD corrections.
Requiring the lepton to be isolated reduces the hard photon
contribution to the ${\cal O}(\alpha^3)$ cross section, and hence 
increases the effect of the
QED corrections. QCD corrections are only slightly modified by
detector effects. 

Since the ${\cal O}(\alpha)$ QED corrections and the ${\cal O}(\alpha_s)$ QCD
corrections are of similar magnitude in the muon case when realistic
experimental conditions are taken into account, one expects that the
${\cal O}(\alpha\alpha_s)$ and ${\cal O}(\alpha_s^2)$ corrections are
also of similar size in this channel. The ${\cal O}(\alpha\alpha_s)$
corrections may thus be non-negligible in a precise determination of the
integrated luminosity from the $Z\to\mu^+\mu^-$ cross section.

\begin{figure}
\phantom{x}
\vskip 15cm
\includegraphics{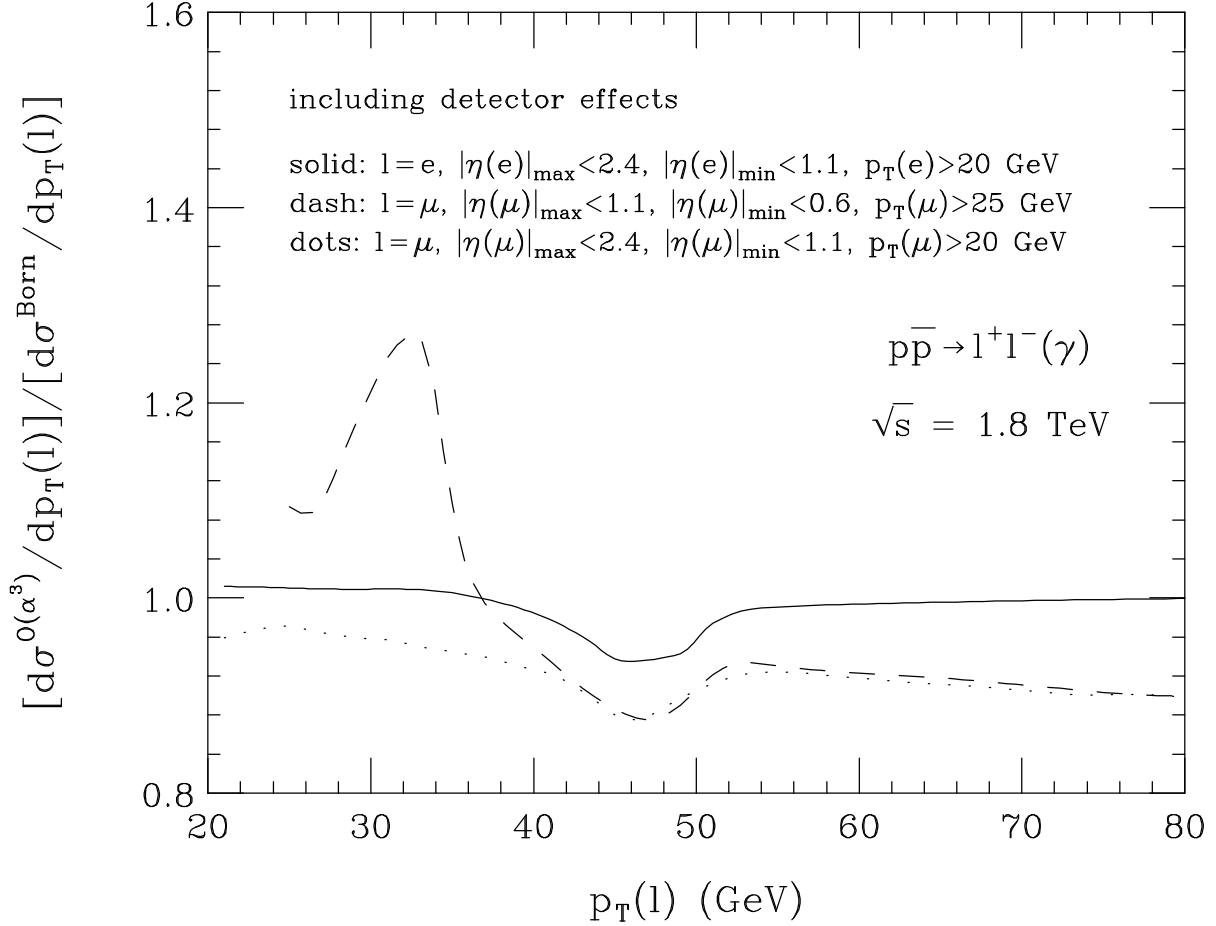}
\caption{Ratio of the \protect{${\cal O}(\alpha^3)$} and lowest order
differential cross sections, including detector effects (see text for
details), as a function of the lepton transverse 
momentum in the reaction $p\bar p\to\ell^+\ell^-(\gamma)$ at 
$\protect{\sqrt{s}=1.8}$~TeV. The solid and dashed lines show the 
cross section ratio for electrons and muons, respectively, employing the
acceptance cuts listed in the text. The dotted line displays the results
for muons if the same pseudorapidity and $p_T$ cuts as for electrons
are used [$|\eta(\ell)|_{\rm max}=\max(|\eta(\ell^+)|,|\eta(\ell^-)|)$; 
$|\eta(\ell)|_{\rm min}=\min(|\eta(\ell^+)|,|\eta(\ell^-)|)$]. }
\label{FIG:ELEVEN}
\end{figure}

Finite detector acceptance cuts do not significantly
modify the QED corrections to $d\sigma/dm(\ell^+\ell^-)$ and $A_{FB}$,
except in the threshold region, $m(\ell^+\ell^-)\approx 
2p_T^{cut}(\ell)$. The effect of the cuts can be more pronounced 
in other distributions. As an example, we show the ratio of the 
lepton transverse momentum distribution at ${\cal O}(\alpha^3)$ and in
the Born approximation in Fig.~\ref{FIG:ELEVEN}. All criteria which are
necessary to simulate lepton detection and identification,
except the isolation cut of Eq.~(\ref{EQ:ISO}), are imposed in this
figure. For the CDF inspired pseudorapidity and $p_T$ cuts we use in
the muon case, the $2\to 2$ phase space becomes much more restricted
than the $2\to 3$ phase space close to the $p_T$ threshold. As a result,
the cross section ratio exhibits a bump located at
$p_T(\mu)\approx 30$~GeV (dashed line). Replacing the acceptance cuts
by those used for electrons, the bump in the cross section disappears
(dotted line). For $p_T(\mu)>40$~GeV, the size of the radiative
corrections is almost independent of the pseudorapidity and transverse 
momentum cuts imposed. Radiative corrections smear out the Jacobian
peak,
causing a characteristic dip in the cross section ratio at $p_T(\mu)
\approx M_Z/2$. However, in this region, the cross section is subject 
to large QCD corrections~\cite{resum} which are not taken into account 
in our calculation. 

The QED corrections to the muon transverse momentum distribution reduce the
cross section by 10~--~15\% over most of the $p_T$ range. For comparison,
we also display the ratio of differential cross sections for $p\bar p\to 
e^+e^-(\gamma)$ in Fig.~\ref{FIG:ELEVEN}. Here, the ${\cal O}(\alpha)$ QED
corrections are of ${\cal O}(1\%)$, except for the Jacobian peak
region, $p_T(e)\approx 45$~GeV, where they reduce the 
cross section by up to~7\%. The pronounced difference in radiative
corrections between electrons and muons is largely due to the different
lepton identification requirements discussed earlier in this subsection.

\subsection{Radiative Corrections and the $Z$ Boson Mass}

As we have seen, final state bremsstrahlung 
severely distorts the Breit-Wigner shape of the $Z$ resonance curve. As
a result, QED corrections must be included when the $Z$ boson mass is
extracted from data, otherwise the mass extracted is shifted to a lower
value. In absence of detector effects, the $Z$ mass shift is 
approximately given by~\cite{Berends}
\begin{equation}
\Delta M_Z\approx -{\pi\beta\over 8}~\Gamma_Z,
\end{equation}
with $\beta$ defined in Eq.~(\ref{EQ:BETA}).
For $Z\to e^+e^-$ ($Z\to\mu^+\mu^-$), $\Delta M_Z\approx -110$~MeV 
($\Delta M_Z\approx -60$~MeV). However, as it is clear from the previous
section, detector effects significantly modify $\Delta M_Z$. 

The $Z$ boson mass extracted from Tevatron experiments 
serves as a reference point when compared with the precise measurement
performed at LEP. It helps to calibrate the electromagnetic energy
scale, and to determine the electron energy resolution as well as the muon
momentum resolution which are important for the measurement of the $W$ mass.

In the approximate treatment of
the QED corrections used so far by the Tevatron experiments, only final
state corrections are taken into account. In addition, the effects of
soft and virtual
corrections are estimated from the inclusive ${\cal O}(\alpha^2)$ 
$Z\to\ell^+\ell^-(\gamma)$ width~\cite{Albert} and the hard photon 
bremsstrahlung contribution~\cite{BK}. 

We now study the differences in the $Z$ boson
masses extracted using the approximation currently employed in the
experimental analysis and our complete ${\cal O}(\alpha^3)$ QED 
calculation, and investigate the effect of the initial state radiative 
corrections on the $Z$ mass shift. To extract the $Z$ boson mass, we use
a log-likelihood fit to the shape of the di-lepton invariant mass 
distribution in the range 81~GeV$<m(\ell^+\ell^-)<101$~GeV. The
templates for the $m(\ell^+\ell^-)$ distributions are calculated using 
the lowest
order differential cross section, varying $M_Z$ between 90.6~GeV and
91.5~GeV in steps of 100~MeV. Detector effects are simulated
as described in Sec.~IIIB. No isolation cut [Eq.~(\ref{EQ:ISO})]
is imposed on the charged leptons. The soft and collinear cutoff 
parameters are chosen to be $\delta_s=10^{-3}$ and 
$\delta_c=3\times 10^{-4}$. In order to be able to determine 
$\Delta M_Z$, it is necessary to properly include the radiation of 
photons with an energy which is of the same order as the shift in $M_Z$,
using the full $2\to 3$ phase space.
$\delta_s$ and $\delta_c$, therefore, have to be smaller than about
$2\times 10^{-3}$, otherwise a 
non-negligible dependence of the $Z$ boson mass shift, $\Delta M_Z$, on 
these parameters remains. 

The error on the 
$Z$ mass resulting from the statistical uncertainties in the Monte
Carlo event samples and the finite step size in varying $M_Z$ in the
templates is approximately 5~MeV in our simulation. This is adequate for 
the semi-quantitative analysis reported here. It is straightforward 
to reduce the uncertainty by increasing the number of events generated and
the number of templates used, given sufficient computing power. 

For definiteness, we concentrate on the electron channel. Results
similar to those which we obtain are expected in the muon case. The shift in
$M_Z$ induced by the QED corrections is determined by comparing the
shape of the ${\cal O}(\alpha^3)$ $e^+e^-$ invariant mass distribution
for the nominal value of $M_Z=91.187$~GeV with that of the templates,
and calculating the log-likelihood as a function of the $Z$ boson mass
used as input in the templates. Repeating this procedure 1000 times with
10,000 events each, the difference between the average of the mass which
maximizes the log-likelihood and the nominal $Z$ boson mass is then
identified with the shift induced by the QED corrections. The same 
procedure is carried out to compute the $Z$ mass shift if the 
approximate calculation of Ref.~\cite{BK} is used. The $Z$ boson mass 
obtained from the complete ${\cal O}(\alpha^3)$ cross section is found 
to be about 10~MeV smaller than that obtained using the approximate 
calculation. Most of the change can be attributed to the different
treatment of the final state soft and virtual corrections in the two
calculations. A change of 10~MeV in $M_Z$ translates into a shift
of several MeV in $M_W$ through the dependence of the energy scale and
the momentum resolution on the $Z$ boson mass 
measured~\cite{CDFWmass,D0Wmass}. For the current level of precision,
this small shift is unimportant. However, it cannot be ignored for a
high-precision measurement of $M_W$. 

In order to estimate how initial state corrections and initial -- final
state interference correction terms affect the $Z$ boson mass, we
compare the mass
obtained using the full ${\cal O}(\alpha)$ corrections with that
extracted when final state radiative corrections are taken into account
only. The two values of $M_Z$ are found to agree within the numerical 
accuracy of our simulation. Initial state radiative corrections and
initial -- final state interference correction terms therefore 
contribute very little to the $Z$ boson mass shift. As we have discussed in 
Sec.~II, current fits to the PDF's do not include QED
effects. This introduces theoretical uncertainties, such as a strong
dependence of the initial state corrections on the factorization scheme
used. 
However, since initial state corrections essentially do not contribute to
the $Z$ boson mass shift, these uncertainties will have no
significant effect on the $Z$ boson mass extracted. This conjecture is
supported by the fact that the numerical values for the mass shifts in the 
QED ${\rm \overline{MS}}$ and DIS scheme are the same.

The $Z$ boson mass extracted from the fit to the di-lepton invariant
mass distribution also depends on the PDF uncertainties, and the choice of
the renormalization and factorization scale. At present, PDF's which
take into account uncertainties in their fit are not generally
available\footnote{An approach to extract PDF's including systematic errors 
has recently been described in Ref.~\cite{Alekhin}.}. 
We therefore only consider the scale dependence here.
Changing $Q^2=\mu^2=M_{QED}^2=M_{QCD}^2$ from $Q^2=\hat s$ to 
$Q^2=100\,\hat s$
decreases the fitted $Z$ mass by 10~MeV both in the Born approximation
and when ${\cal O}(\alpha)$ corrections are taken into account. This
indicates that the $Z$ boson mass shift caused by QED corrections is 
insensitive to the choice of $Q^2$. The scale dependence of the fitted 
$Z$ mass is eliminated when ${\cal O}(\alpha_s)$ QCD corrections are 
taken into account. 

\section{The Forward Backward Asymmetry at the LHC}

As we have mentioned in the Introduction, one can use 
$\sin^2\theta^{lept}_{eff}$
together with $m_{top}$ to constrain the Higgs boson mass. At LEP, 
$\sin^2\theta^{lept}_{eff}$ has been measured with an accuracy of
approximately $\pm 0.00024$~\cite{Renton}. In order to extract the Higgs
boson mass with a precision of $\delta M_H/M_H\approx 30\%$ or better,
the uncertainty in $\sin^2\theta^{lept}_{eff}$ has to be reduced by at
least a factor two. At the LHC ($pp$ collisions at $\sqrt{s}=14$~TeV), the 
$Z\to\ell^+\ell^-$ cross section is approximately 1.6~nb for each lepton
flavor. For the projected yearly integrated luminosity of
100~fb$^{-1}$, this results in a very large number of $Z\to\ell^+\ell^-$
events which, in principle, can be used to measure the forward
backward asymmetry and thus $\sin^2\theta^{lept}_{eff}$ with extremely
high precision~\cite{Fisher}. In this Section, we investigate the
prospects to measure $\sin^2\theta^{lept}_{eff}$ using the forward 
backward asymmetry at the LHC, taking into account the ${\cal
O}(\alpha)$ QED and ${\cal O}(\alpha_s)$ QCD corrections. At LHC 
luminosities, it is easier to trigger on $\mu^+\mu^-$ than on $e^+e^-$ 
pairs in the $Z$ mass region~\cite{ATLAS,CMS}. In our analysis, we 
therefore concentrate on the $Z\to\mu^+\mu^-$ channel;
qualitatively similar results are obtained for the electron channel.

In $pp$ collisions, the quark direction in the initial state has to be
extracted from the boost direction of the di-lepton system with respect 
to the beam axis~\cite{Dittmar}. The cosine of the angle between the 
lepton and the quark in the $\ell^+\ell^-$ rest frame is then
approximated by
\begin{equation}
\cos\theta^*={|p_z(\ell^+\ell^-)|\over p_z(\ell^+\ell^-)}~{2\over 
m(\ell^+\ell^-)\sqrt{m^2(\ell^+\ell^-)
+p_T^2(\ell^+\ell^-)}}\left [p^+(\ell^-)p^-(\ell^+)-p^-(\ell^-)
p^+(\ell^+)\right ].
\label{EQ:CSTAR1}
\end{equation}
For the definition of $\cos\theta^*$ given in Eq.~(\ref{EQ:CSTAR}),
$A_{FB}=0$ for $pp$ collisions.

At the LHC, the sea -- sea quark flux is much larger than at the
Tevatron.
As a result, the probability, $f_q$, that the quark direction and the
boost direction of the di-lepton system coincide is significantly smaller 
than one. The forward backward asymmetry is therefore smaller than at
the Tevatron.
Events with a large rapidity of the di-lepton system, $y(\ell^+\ell^-)$,
originate from collisions where at least one of the partons carries a
large fraction $x$ of the proton momentum. Since valence quarks
dominate at high values of $x$, a cut on the di-lepton rapidity 
increases $f_q$, and thus the asymmetry~\cite{Dittmar} and the
sensitivity to the effective weak mixing angle. 

\begin{figure}
\phantom{x}
\vskip 15cm
\includegraphics{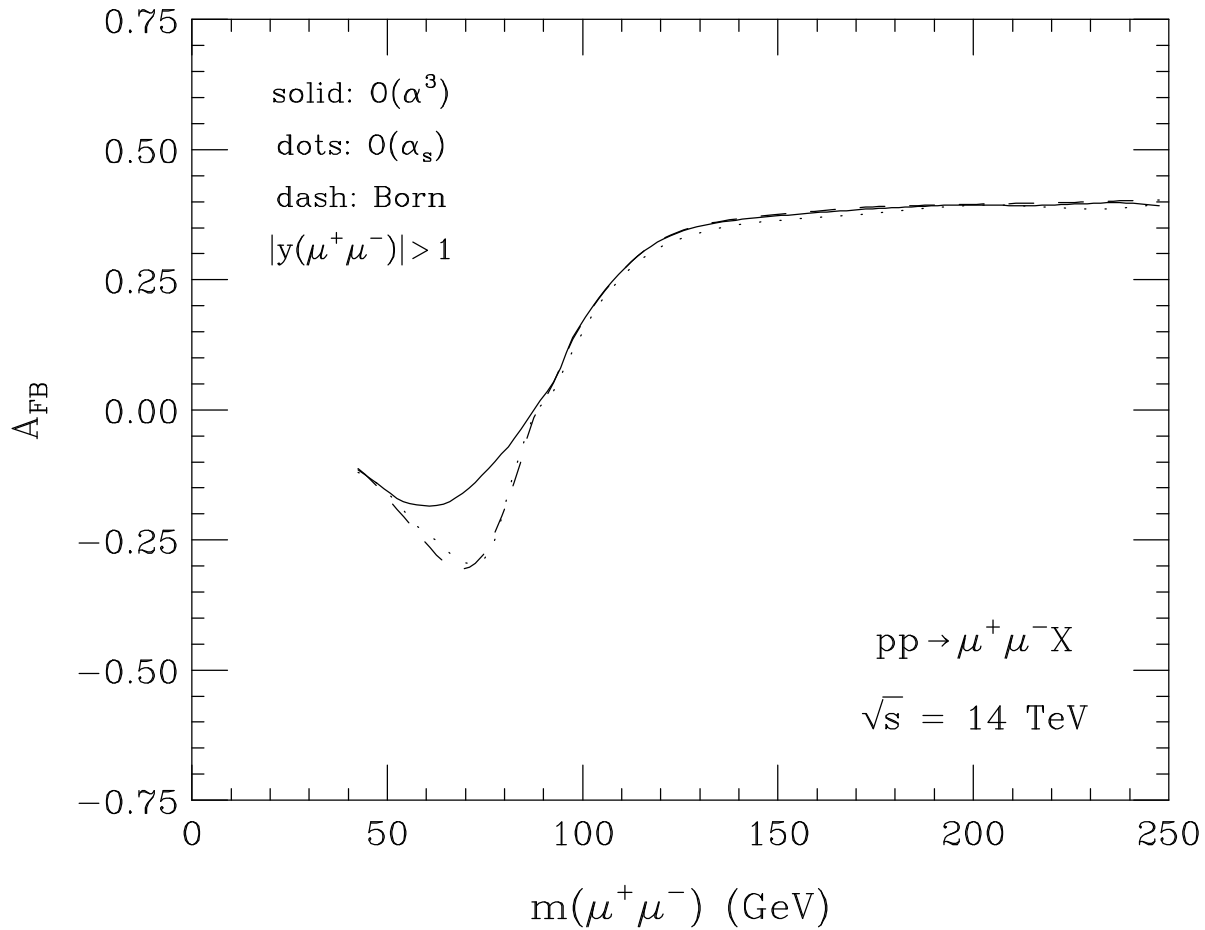}
\caption{The forward backward asymmetry, $A_{FB}$, as a function of the
$\mu^+\mu^-$ invariant mass for $pp\to\mu^+\mu^-(\gamma)$ at
$\protect{\sqrt{s}=14}$~TeV. The solid and dotted lines show the
forward backward asymmetry including ${\cal O}(\alpha)$ QED and ${\cal
O}(\alpha_s)$ QCD corrections, respectively. The dashed line displays the
lowest order prediction of $A_{FB}$. A $|y(\mu^+\mu^-)|>1$ cut is
imposed on the rapidity of the muon pair. No detector effects are
included here.}
\label{FIG:TWELVE}
\end{figure}

The forward backward asymmetry at the LHC, using Eq.~(\ref{EQ:CSTAR1}) 
to define $\cos\theta^*$, and imposing a 
\begin{equation}
|y(\mu^+\mu^-)|>1
\label{EQ:RCUT}
\end{equation}
cut, is shown in Fig.~\ref{FIG:TWELVE} for values of $m(\mu^+\mu^-)$ 
up to 250~GeV. No other cuts besides the $y(\mu^+\mu^-)$ cut have been
imposed in Fig.~\ref{FIG:TWELVE}. Without the cut
of Eq.~(\ref{EQ:RCUT}), $A_{FB}$ would be approximately a factor 1.25
smaller. Although the di-lepton rapidity cut enhances the asymmetry, it
is about a factor~1.5 smaller than at the Tevatron. 

Qualitatively, the behaviour of the forward backward asymmetry as a 
function of the
di-lepton invariant mass is similar to that in $p\bar p$ collisions.
Furthermore, QED and QCD corrections are seen to have a quantitatively 
similar effect on $A_{FB}$ as in $p\bar p$ collisions. In the $Z$ peak 
region, $75~{\rm GeV}<m(\mu^+\mu^-)<105$~GeV, the integrated forward
backward asymmetry can again be 
parameterized by Eqs.~(\ref{EQ:AFB}) and~(\ref{EQ:AFBQED}) with
\begin{equation}
a^{\rm Born}=0.2458, \qquad\qquad b^{\rm Born}=2.19,
\label{EQ:AFBLHC}
\end{equation}
\begin{equation}
\Delta a^{\rm QED}=0.0008, \qquad\qquad \Delta b^{\rm QED}=-0.09.
\label{EQ:AFBLHC2}
\end{equation}
and
\begin{equation}
\Delta a^{\rm QCD}=-0.0011, \qquad\qquad \Delta b^{\rm QCD}=0.06.
\label{EQ:AFBLHC3}
\end{equation}
From Eqs.~(\ref{EQ:AFBLHC}) and~(\ref{EQ:AFBBORN}) we observe that 
the parameter $a$ is essentially the same as at Tevatron energies. $b$,
on the other hand, which controls the sensitivity to the weak mixing angle, 
is significantly reduced. QED corrections increase the integrated
asymmetry in the peak region
by about $\delta A_{FB}/A_{FB}\approx 0.02$, and slightly reduce the
sensitivity to $\sin^2\theta^{lept}_{eff}$. QCD corrections are found to
reduce it by approximately $\delta A_{FB}/A_{FB}\approx -0.
05$. QED and QCD corrections to the integrated forward
backward asymmetry in the $Z$ peak region have opposite sign, as at the
Tevatron.

Using Eqs.~(\ref{EQ:AFBLHC}) and~(\ref{EQ:AFBLHC2}) together with
Eqs.~(\ref{EQ:DEFAFB}) and~(\ref{EQ:DEFFB}), it is now
straightforward to estimate the error expected for
$\sin^2\theta^{lept}_{eff}$ from a measurement of the forward backward
asymmetry in the $Z$ peak region at the LHC. For an integrated
luminosity of 100~fb$^{-1}$, we find that it should be possible to
measure $\sin^2\theta^{lept}_{eff}$ with a statistical precision of
\begin{equation}
\delta\sin^2\theta^{lept}_{eff}=3.9\times 10^{-5}.
\label{EQ:PREC}
\end{equation}
Both, NLO QCD and QED corrections have been taken into account in this
estimate. The $|y(\mu^+\mu^-)|>1$ cut improves the precision for 
$\sin^2\theta^{lept}_{eff}$ by about 10\%. Our result is about 35\% 
better than the estimate given in Ref.~\cite{Fisher}. The shift in $A_{FB}$
introduced by the combined QED and QCD radiative corrections is about 
a factor~7 larger than the expected statistical error.

\begin{figure}
\phantom{x}
\vskip 15cm
\includegraphics{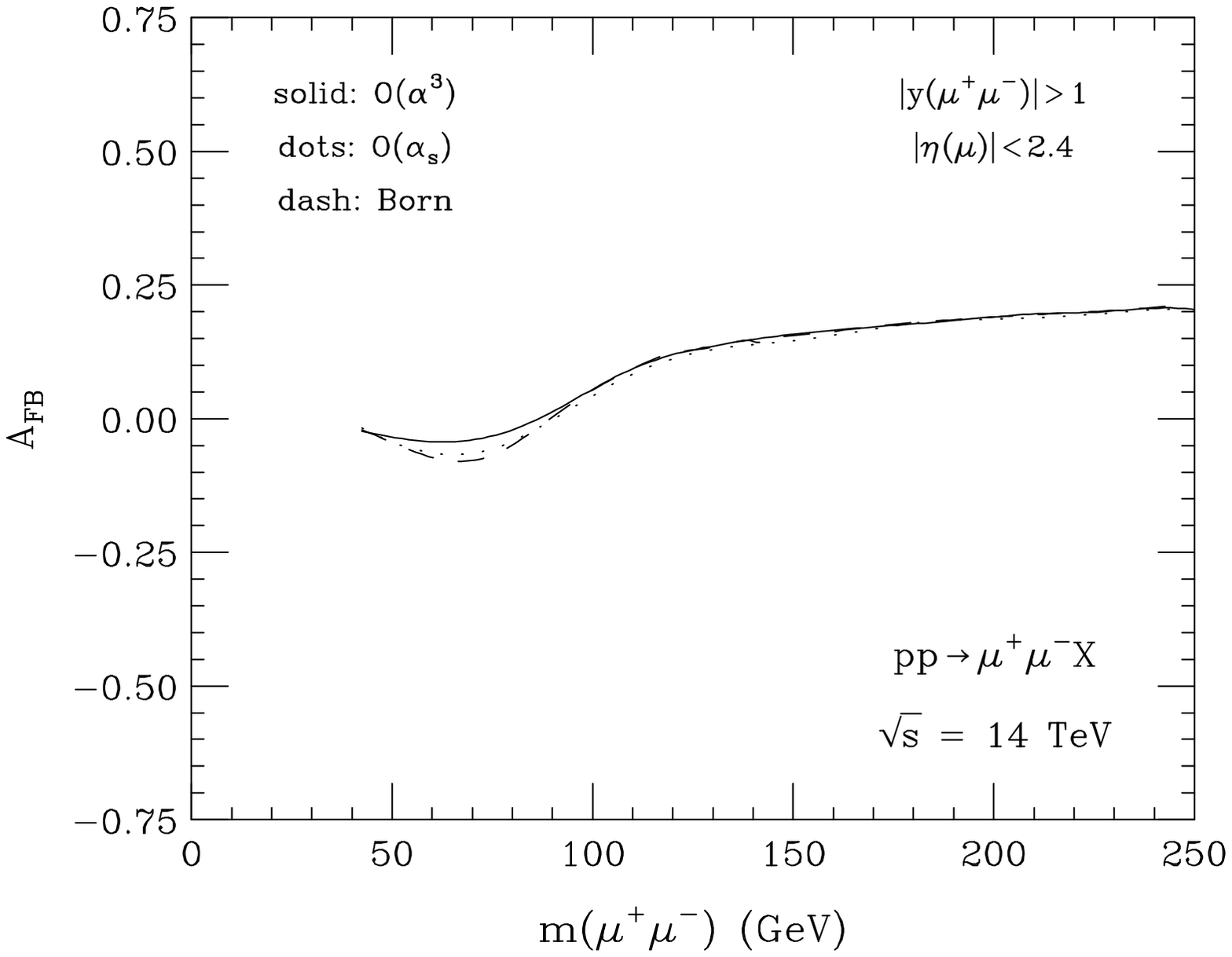}
\caption{The forward backward asymmetry, $A_{FB}$, as a function of the
$\mu^+\mu^-$ invariant mass for $pp\to\mu^+\mu^-(\gamma)$ at
$\protect{\sqrt{s}=14}$~TeV. A $|\eta(\mu)|<2.4$ cut is imposed in
addition to the $|y(\mu^+\mu^-)|>1$ cut. The solid and dotted lines show the
forward backward asymmetry including ${\cal O}(\alpha)$ QED and ${\cal
O}(\alpha_s)$ QCD corrections, respectively. The dashed line displays the
lowest order prediction of $A_{FB}$. }
\label{FIG:THIRTEEN}
\end{figure}

The estimate of Eq.~(\ref{EQ:PREC}) has been obtained assuming full
rapidity coverage for the muons. The proposed FELIX 
experiment~\cite{FELIX} is expected to achieve this. However, FELIX will
operate at a reduced luminosity of at most 
${\cal L}=10^{33}~{\rm cm}^{-2}{\rm s}^{-1}$, corresponding to a yearly
integrated luminosity of 10~fb$^{-1}$ at best. For 10~fb$^{-1}$ the 
expected precision is $\delta\sin^2\theta^{lept}_{eff}\approx 
1.2\times 10^{-4}$. In both the ATLAS and CMS detector, muons can only 
be detected for pseudo-rapidities $|\eta(\mu)|<2.4$~\cite{ATLAS,CMS}. In 
Fig.~\ref{FIG:THIRTEEN} we
display the forward backward asymmetry at the LHC imposing a $|\eta(\mu)
|<2.4$ cut in addition to the di-lepton rapidity cut of 
Eq.~(\ref{EQ:RCUT}). The finite rapidity range covered by the detector
is seen to dramatically reduce the asymmetry. In the region around the
$Z$ pole, the integrated forward backward asymmetry is again an 
approximately linear function of 
$\sin^2\theta^{lept}_{eff}$ (see Eq.~(\ref{EQ:AFB})) with
\begin{equation}
a^{\rm Born}=0.2464, \qquad\qquad b^{\rm Born}=0.72.
\label{EQ:AFBLHC4}
\end{equation}
QED and QCD radiative corrections shift these values by
\begin{equation}
\Delta a^{\rm QED}=0.0024, \qquad\qquad \Delta b^{\rm QED}=-0.07,
\label{EQ:AFBLHC5}
\end{equation}
and
\begin{equation}
\Delta a^{\rm QCD}=0.0067, \qquad\qquad \Delta b^{\rm QED}=-0.27,
\label{EQ:AFBLHC6}
\end{equation}
respectively.
The parameter $b$, which directly controls the sensitivity to 
$\sin^2\theta^{lept}_{eff}$, is reduced by about a factor~3 by
the finite rapidity acceptance. QED corrections further reduce $b$ by
approximately 10\%, and QCD corrections by an additional 30\%. The 
finite rapidity coverage also
results in a reduction of the total $Z$ boson cross section by roughly a
factor~5. As a result, the uncertainty expected for 
$\sin^2\theta^{lept}_{eff}$ with 100~fb$^{-1}$ increases by more than a 
factor~10 to 
\begin{equation}
\delta\sin^2\theta^{lept}_{eff}=4.4\times 10^{-4}\qquad {\rm for} \qquad
|\eta(\mu)|<2.4.
\label{EQ:PREC1}
\end{equation}
The shift in $A_{FB}$ introduced by the combined QCD and QED radiative 
corrections is about a factor~1.5 larger than the statistical error expected.

The rapidity range covered by the electromagnetic calorimeter and the
tracking system is very similar to that of the muon 
system~\cite{ATLAS,CMS}. For $e^+e^-$ production, one therefore does not
expect to measure $\sin^2\theta^{lept}_{eff}$ with a higher
precision than in the muon channel. As in the Tevatron case discussed in
Sec.~IIIA, the QED corrections to $A_{FB}$ in absence of detector
effects are more pronounced in the electron case. A $p_T(\ell)>20$~GeV 
cut has essentially no effect on the forward backward asymmetry in the 
$Z$ peak region. 

The precision expected for $\sin^2\theta^{lept}_{eff}$ from LHC
experiments should be compared with the accuracy from current LEP and
SLC data~\cite{Renton}, and with the sensitivity expected from future
experiments at the SLC and the Tevatron. The combined uncertainty of 
$\sin^2\theta^{lept}_{eff}$ from LEP and SLC experiments is
approximately $2.4\times 10^{-4}$. With the planned luminosity
upgrade~\cite{SLC2000}, one hopes to collect $3\times 10^6$ $Z$ boson 
events at the SLC. This would allow to measure
$\sin^2\theta^{lept}_{eff}$ from the left right asymmetry with a
precision of about $1.2\times 10^{-4}$, which is similar to the one
attainable by FELIX with 10~fb$^{-1}$. At the Tevatron, with the same
integrated luminosity, one expects an uncertainty of $2.3\times
10^{-4}$ for $\sin^2\theta^{lept}_{eff}$~\cite{Tev2000} per experiment. 
In order to improve the precision beyond that expected from future SLC 
and Tevatron experiments, it will be necessary to detect leptons
in the very forward pseudorapidity range, $|\eta|=3.0 - 5.0$
at the LHC when it operates at the design luminosity of
${\cal L}=10^{34}~{\rm cm}^{-2}\,{\rm s}^{-1}$. 

\section{Summary and Conclusions}

In a precision measurement of $M_W$ in hadronic collisions, a 
simultaneous determination of $M_Z$ in di-lepton production is required 
for calibration purposes. The forward backward asymmetry makes it
possible to determine $\sin^2\theta^{lept}_{eff}$ with high precision. 
Both measurements help to constrain the Higgs boson mass from radiative
corrections. In order to perform these high precision measurements, it
is crucial to fully control higher order QCD and electroweak corrections. 
In this paper we have presented a calculation of di-lepton production in
hadronic collisions based on a combination of analytic and
Monte Carlo integration techniques which includes initial and final 
state ${\cal O}(\alpha)$ QED corrections. Previous calculations~\cite{BK,RGW}
have been based on the final state photonic corrections, 
estimating the virtual corrections indirectly from the
inclusive ${\cal O}(\alpha^2)$ $Z\to\ell^+\ell^-(\gamma)$ width and the
hard photon bremsstrahlung contribution. 

\begin{figure}
\phantom{x}
\vskip 15cm
\includegraphics{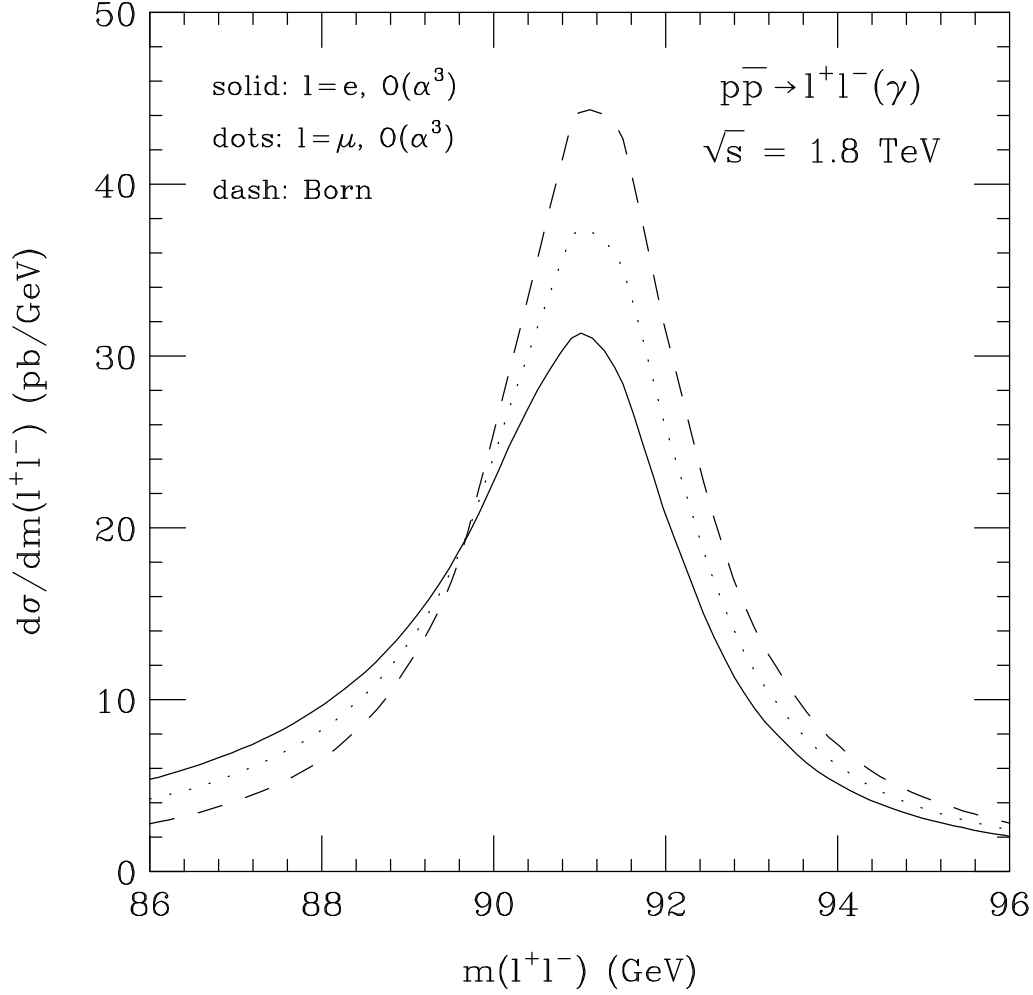}
\caption{The lepton pair invariant mass distribution for $p\bar 
p\to\ell^+\ell^-(\gamma)$ at $\protect{\sqrt{s}=1.8}$~TeV in the 
vicinity of the $Z$ 
peak. The solid (dotted) line shows $d\sigma/dm(\ell^+\ell^-)$ for electron 
(muon) final states including ${\cal O}(\alpha)$ QED corrections. The
dashed lines gives the $\protect{\ell^+\ell^-}$ Born cross section.}
\label{FIG:THREE}
\end{figure}

\begin{figure}
\phantom{x}
\vskip 15cm
\includegraphics{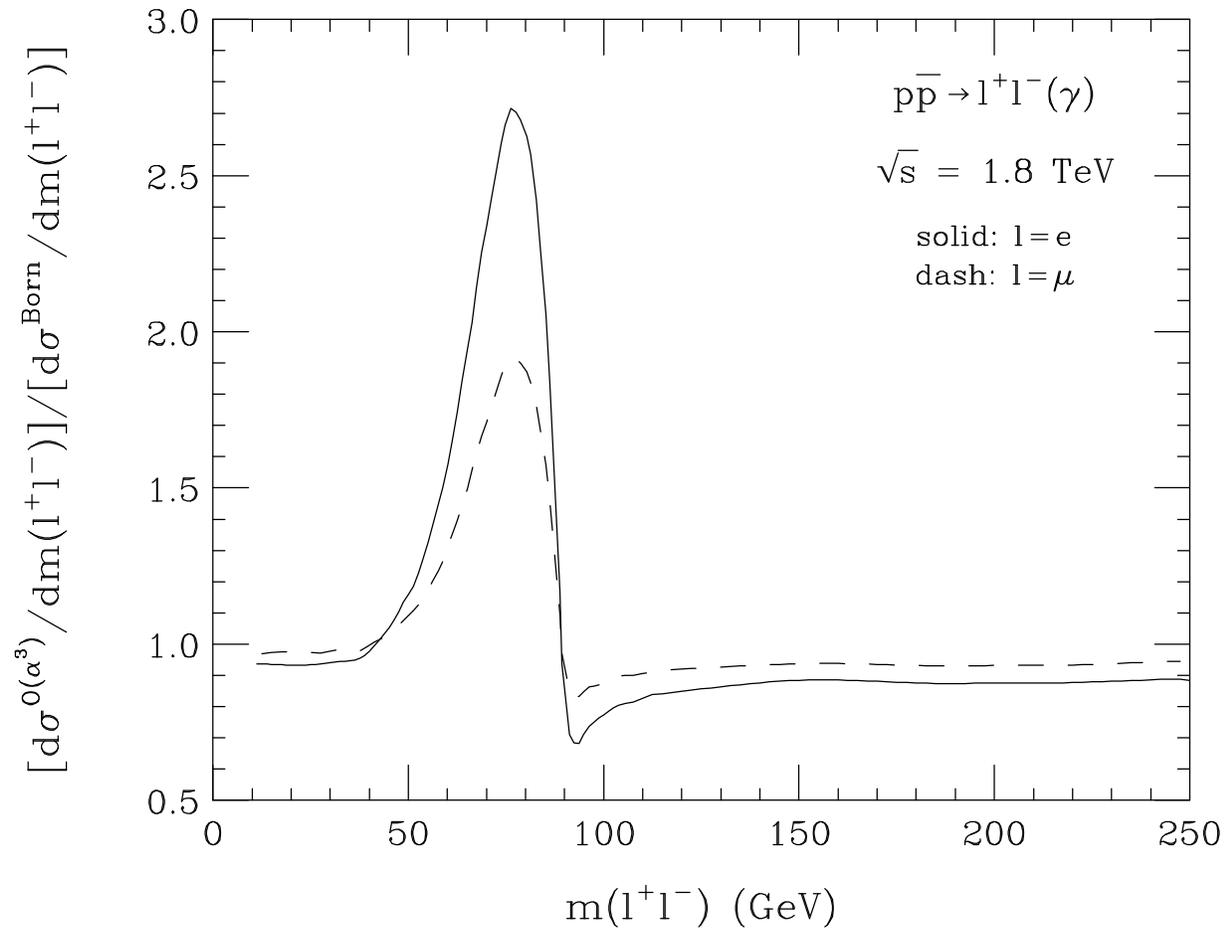}
\caption{Ratio of the \protect{${\cal O}(\alpha^3)$} and lowest order
differential cross sections as a function of the di-lepton invariant
mass for $p\bar p\to\ell^+\ell^-(\gamma)$ at 
$\protect{\sqrt{s}=1.8}$~TeV. The solid line shows the result obtained 
for final state electrons, whereas the dashed line displays the cross 
section ratio for muons.}
\label{FIG:FOUR}
\end{figure}

Due to mass singular logarithmic
terms associated with final state photon radiation in the limit where
the photon is collinear with one of the leptons, final state radiation
effects dominate. Initial state corrections were found to be small
after factorizing the corresponding collinear singularities into the parton
distribution functions. QED corrections to the evolution of the parton 
distribution functions and purely weak corrections are not included in our 
calculation; they are expected to be small. 
Initial state QED corrections are uniform over the entire di-lepton
invariant mass range. In contrast, final state corrections vary rapidly
with $m(\ell^+\ell^-)$, and strongly modify the shape of the invariant
mass distribution as a large fraction of the events shifts from the $Z$
boson peak to lower invariant masses (see Figs.~\ref{FIG:THREE}
and~\ref{FIG:FOUR}). Below $M_Z$, radiative corrections enhance the 
cross section by up to a factor 2.7 (1.9) for electrons (muons).

QED corrections also strongly reduce the magnitude of the forward backward
asymmetry, $A_{FB}$, for di-lepton invariant masses 
between 50~GeV and 90~GeV. In the $Z$ peak region, $75~{\rm 
GeV}<m(\ell^+\ell^-)<105$~GeV, they enhance the integrated
forward backward asymmetry by up to 8\%. 

When detector effects are taken into account, the effect of the mass 
singular logarithmic terms in the electron case is strongly reduced. 
The granularity of the detector and the size of the electromagnetic
showers in the calorimeter make it difficult to 
discriminate between electrons and photons with a small opening
angle. One therefore combines the electron and photon four momentum
vectors if both particles traverse the same calorimeter cell.
In the muon case, the energy of the photon is required to be smaller
than a critical value, $E_c^\gamma$, if both particles traverse the same
calorimeter cell, and  mass singular terms survive. Removing energetic
photons reduces (enhances) the effect of the ${\cal O}(\alpha)$ corrections
below (above) $M_Z$. Detector effects are also
found to considerably decrease the size of the QED corrections to the
forward backward asymmetry below the $Z$ peak for $p\bar 
p\to\mu^+\mu^-(\gamma)$.

QED corrections have a significant impact
on the di-lepton cross section in the $Z$ peak region, and the $Z$ mass 
extracted from experiment. In future Tevatron runs, the total $W/Z$ cross
section may be used as a luminosity monitor~\cite{CDFZSIGN}. As shown
in Table~\ref{TAB:TWO}, QED 
corrections can reduce the di-lepton cross section in the $Z$ peak
region by up to 10\%. 
Final state radiative corrections are known~\cite{CDFWmass,D0Wmass} to
substantially shift the $Z$ boson mass. The $Z$ boson mass extracted 
from our ${\cal O}(\alpha^3)$ $\ell^+\ell^-$ invariant mass distribution
was found to be about 10~MeV smaller than 
that obtained using the approximate calculation of Ref.~\cite{BK}. 
Initial state corrections and initial -- final state interference terms
only marginally influence the amount the $Z$ 
boson mass is shifted. The contribution of the QED corrections to the 
PDF's is expected to be of the size of the initial state radiative 
corrections that are included in our calculation. It is unlikely to be a
limiting factor in the determination of the $Z$ (and $W$) boson mass in
hadronic collisions. 

For the current level of precision, the approximate calculation of 
Ref.~\cite{BK} appears to be adequate. The small difference in the
$Z$ boson mass obtained in the complete ${\cal O}(\alpha^3)$ and the
approximate calculation, however, 
cannot be ignored if one attempts to measure the $W$ mass with high
precision at hadron colliders. This also raises the question of how
strongly multiple final state photon radiation influences the measured $Z$
boson mass. Using the fragmentation function approach, we have shown
that higher order QED corrections non-trivially modify the shape of 
the di-lepton invariant mass distribution.
They may introduce an additional shift of $M_Z$
by ${\cal O}(10$~MeV), and may have a non-negligible impact on the forward
backward asymmetry. So far, only partial calculations 
exist~\cite{Richter}. A more complete understanding of multiple photon
radiation is warranted. 

Finally, we studied the forward backward asymmetry at the LHC. The very 
large number of $Z$ bosons
produced at the LHC offers an opportunity to accurately measure 
$\sin^2\theta^{lept}_{eff}$ from $A_{FB}$. For the forward backward
asymmetry to be non-zero in $pp$ collisions, the scattering angle has
to be defined with respect to the boost direction of the lepton pair 
along the beam axis. Imposing a $|y(\ell^+\ell^-)|>1$ cut reduces the 
fraction of events where the quark direction is misidentified. It 
enhances the asymmetry by a factor~1.25, and thus improves the 
sensitivity to $\sin^2\theta^{lept}_{eff}$ by about 10\%.
With a detector possessing full rapidity coverage for leptons, 
$\sin^2\theta^{lept}_{eff}$ can in principle be measured with a
precision of $\delta\sin^2\theta^{lept}_{eff}=3.9\times 10^{-5}$ if an
integrated luminosity of 100~fb$^{-1}$ is achieved. 
The shift in $A_{FB}$ introduced by QED and QCD radiative corrections is about 
one order of magnitude larger than the statistical error expected. The 
finite lepton rapidity coverage of the ATLAS and CMS detectors strongly 
reduces $A_{FB}$ and the number of $Z$ bosons produced, which results in
an increase of the uncertainty in $\sin^2\theta^{lept}_{eff}$ by
about a factor~10. In order to significantly improve the precision for
$\sin^2\theta^{lept}_{eff}$ beyond that expected from future SLC 
and Tevatron experiments, it will thus be necessary to detect electrons
and muons in the very forward pseudorapidity range, $|\eta|=3.0 - 5.0$,
at the LHC, and to achieve an integrated luminosity of ${\cal
O}(100~{\rm fb}^{-1})$.

%
\acknowledgements

We would like to thank I.~Adam, A.~Bodek, M.~Demarteau, S.~Errede, 
E.~Flattum, H.~Frisch, F.~Halzen, Y.K.~Kim, E.~Laenen, M.~Strikman, 
C.~Taylor, and D.~Wackeroth for useful and stimulating
discussions. One of us (U.B.) is grateful to the Fermilab Theory Group,
where part of this work was carried out, for its generous hospitality.
This work has been supported in part by Department of Energy 
contract No.~DE-AC02-76CHO3000 and NSF grant PHY-9600770. 

%
%

%
\newpage
%
\widetext
\begin{table}
\caption{The integrated forward backward asymmetry, $A_{FB}$, in 
\protect{$p\bar p\to
e^+e^-X$} at $\protect{\sqrt{s}=1.8}$~TeV for \protect{$75~{\rm 
GeV}<m(e^+e^-)<105$}~GeV and $\protect{m(e^+e^-)>105}$~GeV. Shown are 
the SM predictions with and without \protect{${\cal O}(\alpha)$} QED 
corrections together with the
experimental values of Ref.~[\ref{CDFAFB}]. The uncertainties listed
for the theoretical results represent the statistical error of the Monte 
Carlo integration.\protect{\\}} 
\label{TAB:ONE}
\begin{tabular}{ccc}
\multicolumn{1}{c}{} &
\multicolumn{1}{c}{$75~{\rm GeV}<m(e^+e^-)<105$~GeV} &
\multicolumn{1}{c}{$m(e^+e^-)>105$~GeV} \\
\tableline
$A_{FB}^{Born}$ & $0.048\pm 0.001$ & $0.523\pm 0.001$ \\
$A_{FB}^{{\cal O}(\alpha^3)}$ & $0.052\pm 0.001$ & $0.528\pm 0.001$ \\
$A_{FB}^{exp.}$ & $0.070\pm 0.016$ & $0.43\pm 0.10$\\
\end{tabular}
\end{table}
\vskip 5.mm
\begin{table}
\caption{The cross section ratios $K^{QED}=\sigma^{{\cal O}(\alpha^3)}
/\sigma^{\rm Born}$ and $K^{QCD}=\sigma^{{\cal O}(\alpha_s)}/\sigma^{\rm 
Born}$ for $p\bar p\to\ell^+\ell^-X$ ($\ell=e,\,\mu$) at 
$\protect{\sqrt{s}=1.8}$~TeV
with \protect{$75~{\rm GeV}<m(\ell^+\ell^-)<105$}~GeV. Shown are the 
predictions for three cases: without taking any detector effects
into account (``no detector effects''), with the detector effects 
described in the text and no lepton isolation cut (``with detector
effects, no lepton isolation''), and finally adding lepton isolation 
[see Eq.~(\protect{\ref{EQ:ISO}})] (``with detector effects, with lepton 
isolation'').\protect{\\}}
\label{TAB:TWO}
\begin{tabular}{cccc}
\multicolumn{1}{c}{} &
\multicolumn{1}{c}{no detector effects} &
\multicolumn{2}{c}{with detector effects} \\
\multicolumn{1}{c}{} &
\multicolumn{1}{c}{} &
\multicolumn{1}{c}{no lepton isolation} &
\multicolumn{1}{c}{with lepton isolation} \\
\tableline
$K^{QED}~(p\bar p\to e^+e^-X)$ & 0.93 & 0.98 & 0.96 \\
$K^{QED}~(p\bar p\to \mu^+\mu^-X)$ & 0.97 & 0.92 & 0.90 \\
$K^{QCD}~(p\bar p\to \ell^+\ell^-X)$ & 1.17 & 1.16 & 1.14\\
\end{tabular}
\end{table}

\newpage
%
%
\newpage
%
\newpage
%
\newpage
%
\newpage
%
\newpage
%
\newpage
%
\newpage
%
\newpage
%
\newpage
%
\newpage
%
\newpage
%
\newpage
%
\newpage
%
%
%

\begin{references}
%
\bibitem{Renton}
M.~Demarteau, FERMILAB-Conf/96-354, to appear in the Proceedings of the 
{\sl ``DPF96 Conference''}, Minneapolis, MN, August~10 --~15, 1996; 
D.~Abbaneo {\it et al.} (LEP Electroweak Working Group), CERN-PPE/96-183
(preprint, December~1996).
%
\bibitem{Lanc}
M.~Lancaster, FERMILAB-Conf-97/176-E, to appear in the proceedings of
the {\sl ``Le Recontres de Physique de la Vall\'ee d' Aoste''}, La
Thuile, Italy, March~3 --~7, 1997.
%
\bibitem{tom}
T.~Rizzo, Mod. Phys. Lett. {\bf A6}, 1961 (1991); J.~Ellis, G.~Fogli and
E.~Lisi, Phys. Lett. {\bf B274}, 456 (1992) and Phys. Lett. {\bf B318}, 
148 (1993).
%
\bibitem{Tev2000}
H.~Aihara {\it et al.}, in {\sl ``Future Electroweak Physics at the Fermilab
Tevatron: Report of the $TEV\_2000$ Study Group''}, eds. D.~Amidei and 
R.~Brock, FERMILAB-Pub/96-082, p.~63 (April~1996); U.~Baur and 
M.~Demarteau, FERMILAB-Conf/96-423, to appear in the Proceedings of the
Workshop {\sl ``New Directions in High Energy Physics''}, Snowmass, CO,
June~25 -- July~12, 1996; K.~Hagiwara, D.~Haidt and S.~Matsumoto,
KEK-TH-512, DESY 96-192, (preprint), to appear in Z.~Phys.~{\bf C}.
%
\bibitem{LEPWmass}
A.~Ballestrero {\it et al.}, in {\sl ``Physics at LEP2''}, eds. 
G.~Altarelli, T.~Sjostrand and F.~Zwirner, CERN Yellow Report, 
CERN-96-01, Vol.~1, p.~141.
%
\bibitem{GPJ}
J.P.~Marriner, FERMILAB-Conf-96/391, to appear in the Proceedings of the
Workshop {\sl ``New Directions in High Energy Physics''}, Snowmass, CO,
June~25 -- July~12, 1996;
P.P.~Bagley {\it et al.}, FERMILAB-Conf-96/392, to appear in the 
Proceedings of the
Workshop {\sl ``New Directions in High Energy Physics''}, Snowmass, CO,
June~25 -- July~12, 1996;
D.A.~Finley, J.~Marriner and N.V.~Mokhov, FERMILAB-Conf-96/408,
presented at the {\sl ``Conference on Charged Particle Accelerators''},
Protvino, Russia, October~22 --~24, 1996.
%
\bibitem{KW}
S.~Keller and J.~Womersley, FERMILAB-Conf-96/422-T, to appear in the 
Proceedings of the
Workshop {\sl ``New Directions in High Energy Physics''}, Snowmass, CO,
June~25 -- July~12, 1996.
%
\bibitem{CDFWmass}
F.~Abe {\it et al.} (CDF Collaboration), Phys. Rev. Lett. {\bf 75}, 11
(1995) and Phys. Rev. {\bf D52}, 4784 (1995).
%
\bibitem{D0Wmass}
S.~Abachi {\it et al.} (D\O\ Collaboration), Phys. Rev. Lett. {\bf 77},
3309 (1996), and FERMILAB-Conf/96-251-E, submitted
to the {\sl ``28$^{\rm th}$ International Conference on High Energy
Physics''}, Warsaw, Poland, 25~-- 31~July~1996.
%
\bibitem{alberto}
D.~Bardin {\it et al.}, CERN-TH. 6443/92 (preprint); P.~Gambino and 
A.~Sirlin, Phys. Rev. {\bf D49}, 1160 (1994).
%
\bibitem{Fisher}
P.~Fisher, U.~Becker and P.~Kirkby, Phys. Lett. {\bf B356}, 404 (1995).
%
\bibitem{AFBCDF}
C.~Albajar {\it et al.} (UA1 Collaboration), Z.~Phys. {\bf C44}, 15
(1989);
F.~Abe {\it et al.} (CDF Collaboration), Phys. Rev. Lett. {\bf 67}, 1502
(1991); P.~Hurst (CDF Collaboration), Ph.D. Thesis, University of 
Illinois at Urbana -- Champaign, 1990.
%
\bibitem{BK}
F.~Berends and R.K.~Kleiss, Z.~Phys. {\bf C27}, 365 (1985).
%
\bibitem{RGW}
R.G.~Wagner, Comput. Phys. Commun. {\bf 70}, 15 (1992).
%
\bibitem{CDFDY}
F.~Abe {\it et al.} (CDF Collaboration), Phys. Rev. {\bf D49}, R1
(1994).
%
\bibitem{Rosner}
J.~Rosner, Phys. Lett. {\bf B221}, 85 (1989), and Phys. Rev. {\bf D54},
1078 (1996).
%
\bibitem{CDFZp}
F.~Abe {\it et al.} (CDF Collaboration), Phys. Rev. {\bf D51}, 949
(1995). 
%
\bibitem{CDFAFB}
\label{CDFAFB}
F.~Abe {\it et al.} (CDF Collaboration), Phys. Rev. Lett. {\bf 77}, 2616
(1996).
%
\bibitem{NLOMC}
H.~Baer, J.~Ohnemus, and J.F.~Owens, Phys. Rev. {\bf D40}, 2844 (1989) and 
Phys. Rev. {\bf D42}, 61 (1990); L.~Bergmann, Ph.D. dissertation, 
Florida State University, report No. FSU-HEP-890215, 1989 (unpublished);
W.~Giele and E.W.N.~Glover, Phys. Rev. {\bf D46}, 1980 (1992).
%
\bibitem{BaBe}
U.~Baur and E.L.~Berger, Phys. Rev. {\bf D47}, 4889 (1993).
%
\bibitem{BaZe}
U.~Baur and D.~Zeppenfeld, Phys. Rev. Lett. {\bf 75}, 1002 (1995).
%
\bibitem{spies}
J.~Kripfganz and H.~Perlt, Z.~Phys. {\bf C41}, 319 (1988);
H.~Spiesberger, Phys. Rev. {\bf D52}, 4936 (1995).
%
\bibitem{RPS}
A.~de Rujula, R.~Petronzio and A.~Savoy-Navarro, Nucl. Phys. {\bf B154},
394 (1979).
%
\bibitem{ward}
D.B.~DeLaney {\it et al.}, Phys. Rev. {\bf D47}, 853 (1993); Phys. Lett.
{\bf B292}, 413 (1992), (E) Phys. Lett. {\bf B302}, 540 (1993); 
G.~Siopsis {\it et al.}, Acta Phys. Polon. {\bf B23}, 1133 (1992).
%
\bibitem{LHC}
D.~Bousard {\it et al.} (The LHC Study Group), CERN/AC/95-05
(October~1995).
%
\bibitem{DIMREG}
G.'t Hooft and M.~Veltman,
Nucl. Phys. {\bf B44}, 189 (1972).  
%
\bibitem{AL}
V.N.~Gribov and L.N.~Lipatov, Sov.~J. Nucl. Phys. {\bf 15}, 78 (1972);
Yu.L.~Dokshitzer, JETP {\bf 73}, 1216 (1977);
G.~Altarelli and G.~Parisi, Nucl. Phys. {\bf B126}, 298 (1977).
%
\bibitem{PDF}
A.~Martin, R.G.~Roberts, and W.J.~Stirling, Phys. Lett. {\bf B354}, 155
(1995) and Phys. Lett. {\bf B387}, 419 (1996);
H.~Lai {\it et al.} (CTEQ Collaboration), Phys. Rev. {\bf D51},
4763 (1995) and Phys. Rev. {\bf D55}, 1280 (1997).
%
\bibitem{MSBAR}
W.A.~Bardeen, A.J.~Buras, D.W.~Duke, and T.~Muta,
Phys. Rev. {\bf D18}, 3998 (1978).	
%
\bibitem{OWENSTUNG}J.~F.~Owens and W.~K.~Tung, 
Annu. Rev. Nucl. Part. Sci. {\bf 42}, 291 (1992).
%
\bibitem{BR}
H.~Baer and M.H.~Reno, Phys. Rev. {\bf D43}, 2892 (1991).
%
\bibitem{Hollik}
M.~B\"ohm and W.~Hollik, Nucl. Phys. {\bf B204}, 45 (1982). 
%
\bibitem{HLK}
W.~Hollik, Fort. Phys. {\bf 38}, 165 (1990).
%
\bibitem{MRSA}
A.D.~Martin, R.G.~Roberts, and W.J.~Stirling, 
Phys. Rev. {\bf D50}, 6734 (1994).	
%
\bibitem{KLN}
T. Kinoshita, J.~Math. Phys. {\bf 3}, 650 (1962); T.D.~Lee and 
M.~Nauenberg, Phys. Rev. {\bf 133}, B1549 (1964).
%
\bibitem{Berends}
F.~Berends, in {\sl ``Physics at LEP~1''}, eds. G.~Altarelli, R.~Kleiss
and C.~Verzegnassi, CERN 89-08, Vol.~1, p.~89.
%
\bibitem{NT}
O.~Nicrosini and L.~Trentadue, Z.~Phys. {\bf C39}, 479 (1988).
%
\bibitem{Bardin}
D.~Bardin {\it et al.}, Nucl. Phys. {\bf B351}, 1 (1991).
%
\bibitem{CS}
J.~Collins and D.~Soper, Phys. Rev. {\bf D16}, 2219 (1977).
%
\bibitem{BH1}
M.~B\"ohm and W.~Hollik, Phys. Lett. {\bf B139}, 213 (1984).
%
\bibitem{RCDF}
F.~Abe {\it et~al.} (CDF Collaboration), Phys. Rev. {\bf D45}, 3921
(1992).
%
\bibitem{BR1}
H.~Baer and M.H.~Reno, Phys. Rev. {\bf D45}, 1503 (1992).
%
\bibitem{CALLA}
D.~Callaway, Phys. Lett. {\bf B108}, 421 (1982).
%
\bibitem{D0ZSIG}
S.~Abachi {\it et al.} (D\O\ Collaboration), Phys. Rev. Lett. {\bf 75},
1456 (1995).
%
\bibitem{CDFZSIG}
F.~Abe {\it et~al.} (CDF Collaboration), Phys. Rev. {\bf D44}, 29
(1991) and Phys. Rev. Lett. {\bf 69}, 28 (1992).
%
\bibitem{CDFZSIGN}
F.~Abe {\it et~al.} (CDF Collaboration), Phys. Rev. Lett. {\bf 76}, 3070
(1996).
%
\bibitem{DITT1}
M.~Dittmar, F.~Pauss and D.~Z\"urcher, ETHZ-IPP PR-97-01 (preprint,
May~1997), submitted to Phys. Rev.~{\bf D}.
%
\bibitem{resum}
G.~Altarelli, R.K.~Ellis, M.~Greco and G.~Martinelli, Nucl. Phys. {\bf
B246}, 12 (1984);
P.~Arnold and R.~Kauffman, Nucl. Phys. {\bf B349}, 381 (1991);
H.~Contapanagos and G.~Sterman, Nucl. Phys. {\bf B400}, 211 (1993);
L.~Alvero and H.~Contapanagos, Nucl. Phys. {\bf B456}, 497 (1995);
W.~Giele and S.~Keller, FERMILAB-Pub-96/332-T (preprint, April~1997).
%
\bibitem{Albert}
D.~Albert, W.J.~Marciano, D.~Wyler, and Z.~Parsa, Nucl. Phys. {\bf
B166}, 460 (1980).
%
\bibitem{Alekhin}
S.~Alekhin, hep-ph/9611213 (preprint, November 1996).
%
\bibitem{ATLAS} 
D.~Gingrich {\it et al.} (ATLAS Collaboration), ATLAS Letter of
Intent, CERN-LHCC-92-4 (October~1992); W.~W.~Armstrong {\it et al.}
(ATLAS Collaboration), ATLAS Technical Design Report, CERN-LHCC-94-43 
(December 1994).
%
\bibitem{CMS} 
M.~Della Negra {\it et al.} (CMS Collaboration), CMS Letter of
Intent, CERN-LHCC-92-3 (October~1992); G.~L.~Bayatian {\it et al.}
(CMS Collaboration), CMS Technical Design Report, CERN-LHCC-94-38 
(December 1994).
%
\bibitem{Dittmar}
M.~Dittmar, Phys. Rev. {\bf D55}, 161 (1997).
%
\bibitem{FELIX}
K.~Eggert and C.~Taylor, CERN-PPE/96-136 (preprint, October 1996),
submitted to Nucl. Phys.
%
\bibitem{SLC2000}
M.~Breidenbach {\it et al.}, SLAC-CN-409 (1996).
%
\bibitem{Richter}
U.~Baur, T.~Han, N.~Kauer, R.~Sobey and D.~Zeppenfeld, Phys. Rev. {\bf
D56}, 140 (1997);
E.~Richter-Was, Z.~Phys. {\bf C64}, 227 (1994); S.~Laporta and 
R.~Odorico, Nucl. Phys. {\bf B266}, 633 (1986) and Comp. Phys. Comm.
{\bf 39}, 127 (1986); R.~Odorico, Comp. Phys. Comm. {\bf 59}, 527
(1990).
%
\end{references}
\end{document}